\catcode`\@=11

\newif\if@fewtab\@fewtabtrue

{\count255=\time\divide\count255 by 60
\xdef\hourmin{\number\count255}
\multiply\count255 by-60\advance\count255 by\time
\xdef\hourmin{\hourmin:\ifnum\count255<10 0\fi\the\count255}}
\def\ps@draft{\let\@mkboth\@gobbletwo
    \def\@oddhead{}
    \def\@oddfoot
       {\hbox to 7 cm{$\scriptstyle Draft\ version:\ \draftdate$
       \hfil}\hskip -7cm\hfil\rm\thepage \hfil}
    \def\@evenhead{}\let\@evenfoot\@oddfoot}

\def\ceqno{\global\@fewtabfalse
    \ifcase\@eqcnt \def\@tempa{& & &}\or \def\@tempa{& &}
      \or \def\@tempa{&}
      \or\def\@tempa{}\fi\@tempa
{\rm(\theequation)}}

\def\aeqno#1{\global\@fewtabfalse
    \ifcase\@eqcnt \def\@tempa{& & &}\or \def\@tempa{& &}
      \or \def\@tempa{&}
      \or\def\@tempa{}\fi\@tempa
{\rm(\theequation,#1)}}

\def\label#1{\ifnum\draftcontrol=1
 \global\def\draftnote{$\scriptstyle #1$}\fi
 \@bsphack\if@filesw {\let\thepage\relax
   \def\protect{\noexpand\noexpand\noexpand}%
\xdef\@gtempa{\write\@auxout{\string
      \newlabel{#1}{{\@currentlabel}{\thepage}}}}}\@gtempa
   \if@nobreak \ifvmode\nobreak\fi\fi\fi
  \@esphack}

\def\alabel#1#2{\label{#1}\global\@fewtabfalse
    \ifcase\@eqcnt \def\@tempa{& & &}\or \def\@tempa{& &}
      \or \def\@tempa{&}
      \or\def\@tempa{}\fi\@tempa
{\hbox to 3cm{\phantom{\rm(\theequation,#2)}
\draftnote \hfil}\hskip -3cm {\rm(\theequation,#2)}}}

\def\clabel#1{\label{#1}\global\@fewtabfalse
    \ifcase\@eqcnt \def\@tempa{& & &}\or \def\@tempa{& &}
      \or \def\@tempa{&}
      \or\def\@tempa{}\fi\@tempa
{\hbox to 3cm{\phantom{\rm(\theequation)}
\draftnote \hfil}\hskip -3cm{\rm(\theequation)}}}

\def\eqnarray{\def\draftnote{{}}\global\@fewtabtrue
\stepcounter{equation}\let\@currentlabel=\theequation
\global\@eqnswtrue
\global\@eqcnt\z@\tabskip\@centering\let\\=\@eqncr
$$\halign to \displaywidth\bgroup\@eqnsel\hskip\@centering\@eqcnt\z@
  $\displaystyle\tabskip\z@{##}$&\global\@eqcnt\@ne
  \hskip 1\arraycolsep \hfil${##}$\hfil
  &\global\@eqcnt\tw@ \hskip 1\arraycolsep
$\displaystyle\tabskip\z@{##}$
\hfil  \tabskip\@centering&\global\@eqcnt\thr@@\llap{##}\tabskip\z@
\cr}

\def\endeqnarray{\@@eqncr\egroup
      \global\advance\c@equation\m@ne$$\global\@ignoretrue}

\def\@eqnnum{\hbox to 3cm{\phantom{\rm(\theequation)} \draftnote
                         \hfil}\hskip -3cm {\rm(\theequation)}}

\def\@@eqncr{\let\@tempa\relax
    \ifcase\@eqcnt \def\@tempa{& & &}\or \def\@tempa{& &}
      \or \def\@tempa{&}
      \or\def\@tempa{}
\fi\@tempa
\if@eqnsw
\if@fewtab\@eqnnum\fi
\stepcounter{equation}\fi\global
\@eqnswtrue\global\@eqcnt\z@\global\@fewtabtrue\cr}

\def\draftcite#1{\ifnum\draftcontrol=1#1\else{}\fi}

\def\@lbibitem[#1]#2{\item{}\hskip -3cm \hbox to 2cm
{\hfil$\scriptstyle\draftcite{#2}$}\hskip
1cm[\@biblabel{#1}]\if@filesw
     {\def\protect##1{\string ##1\space}\immediate
      \write\@auxout{\string\bibcite{#2}{#1}}}\fi\ignorespaces}

\def\@bibitem#1{\item\hskip -3cm \hbox to 2cm
{\hfil $\scriptstyle\draftcite{#1}$}\hskip 1cm
\if@filesw \immediate\write\@auxout
       {\string\bibcite{#1}{\the\value{\@listctr}}}\fi\ignorespaces}

\def\nsection#1{\section{#1}\setcounter{equation}{0}}

\font\tendl=msbm10  scaled \magstep1
\font\sevendl=msbm7 scaled \magstep1
\font\fivedl=msbm5 scaled \magstep1
\font\tengl=eufm10  scaled \magstep1
\font\sevengl=eufm7 scaled \magstep1
\font\fivegl=eufm5 scaled \magstep1

\newfam\dlfam  
\textfont\dlfam=\tendl \scriptfont\dlfam=\sevendl
\scriptscriptfont\dlfam=\fivedl
\newfam\glfam  
\textfont\glfam=\tengl \scriptfont\glfam=\sevengl
\scriptscriptfont\glfam=\fivegl

\def\draftdate{\number\month/\number\day/\number\year\ \ \ \hourmin }

\global\def\draftcontrol{0}
\catcode`\@=12
\def\tilde{\widetilde}
\def\hat{\widehat}
\documentclass[aps,amsfonts,amsmath,amssymb,floatfix,nofootinbib]{revtex4}
\usepackage{epsfig}
\usepackage{rotating}
\usepackage[centerlast]{subfigure}
\usepackage{bm}
\usepackage{amssymb}
\usepackage{comment}
\renewcommand{\theequation}{\thesection.\arabic{equation}}
\newcommand{\ii}{\mathrm{i}}
\newcommand{\be}{\begin{eqnarray}}
\newcommand{\en}{\end{eqnarray}\vs 0.5 cm}

\newcommand{\Id}{{I\hspace{-0.04cm}d}}

\newcommand{\vs}{\vskip}

\newcommand{\BZ}{{B\hspace{-0.05cm}Z}}

\newcommand{\qq}{\begin{eqnarray}}

\newcommand{\ee}{{\rm e}}

\newcommand{\qqq}{\end{eqnarray}}

\newcommand{\tr}{\hbox{tr}}

\newcommand{\CC}{{\cal C}}

\newcommand{\CE}{{\cal E}}
\newcommand{\CF}{{\cal F}}
\newcommand{\CG}{{\cal G}}
\newcommand{\CH}{{\cal H}}

\newcommand{\CK}{{\cal K}}
\newcommand{\CL}{{\cal L}}

\newcommand{\CN}{{\cal N}}

\newcommand{\CQ}{{\cal Q}}

\newcommand{\WZ}{{W\hspace{-0.05cm}Z}}
\newcommand{\HD}{{H\hspace{-0.05cm}D}}
\newcommand{\KM}{{K\hspace{-0.05cm}M}}
\begin{document}
\title{{\Large\bf{Bundle gerbes for topological insulators\footnote{Expanded 
version of a mini-course given at the Section of Mathematics, 
University of Geneva, October 6 and 8, 2015.}}}}
\author{Krzysztof Gaw\c{e}dzki}
\affiliation{Laboratoire de Physique, 
ENS Lyon, Universit\'e de Lyon,\\ 46 All\'ee d'Italie, 69364 Lyon, France\\
C.N.R.S., chercheur \'em\'erite}

\date{\today}

\maketitle

\vskip 0.6cm

\centerline{\bf Abstract}
\vskip 0.4cm

\qquad\parbox[t]{15cm}{Bundle gerbes are simple examples 
of higher geometric structures that show their utility in 
dealing with topological subtleties of physical theories. 
I will review a recent construction of torsion topological 
invariants for condensed matter systems via equivariant 
bundle gerbes. The construction covers static and periodically 
driven systems with time reversal invariance in 2 and 3 space 
dimensions. It involves refinements of geometry of gerbes that 
will be discussed in the first lecture, the second one being 
devoted to the applications to topological insulators.}
\vskip 1.5cm

\centerline{\bf Contents}
\vskip 0.7cm

\noindent{\bf Lecture 1. \,Wess-Zumino amplitudes and bundle gerbes}
\vskip 0.4cm

\noindent$\bullet$ Introduction
\vskip 0.1cm

\noindent$\bullet$ $2d$ Wess-Zumino amplitudes and their square root
\vskip 0.1cm

\noindent$\bullet$ Bundle gerbes with connection
\vskip 0.1cm

\noindent$\bullet$ Bundle gerbe holonomy 
\vskip 0.1cm

\noindent$\bullet$ Gerbes equivariant under involution 
\vskip 0.1cm

\noindent$\bullet$ Square root of gerbe holonomy
\vskip 0.1cm

\noindent$\bullet$ $3d$ index with values in $\,\{\pm1\}$
\vskip 0.1cm

\noindent$\bullet$ Basic gerbe over unitary group
\vskip 0.1cm

\noindent$\bullet$ Towards time-reversal equivariant basic
gerbe over $U(N)$
\vskip 0.1cm

\noindent$\bullet$ Lift to the double cover of $\,U(N)$
\vskip 0.7cm

\noindent{\bf Lecture 2. \,Applications to topological insulators}
\vskip 0.4cm

\noindent$\bullet$ Crystalline systems and Bloch theory
\vskip 0.1cm

\noindent$\bullet$ Chern insulators and their homotopic invariant
\vskip 0.1cm

\noindent$\bullet$ Time reversal symmetry and the $2d$ Kane-Mele invariant
\vskip 0.1cm

\noindent$\bullet$ Kane-Mele invariant as the square root of a WZ amplitude
\vskip 0.1cm

\noindent$\bullet$ $3d$ \,Kane-Mele invariants
\vskip 0.1cm

\noindent$\bullet$ Floquet theory of periodically driven crystalline systems
\vskip 0.1cm

\noindent$\bullet$ Topological invariants of gapped Floquet $2d$ and
$3d$ systems
\vskip 0.1cm

\noindent$\bullet$ Bulk-edge correspondence
\vskip 0.1cm

\noindent$\bullet$ Conclusions
\eject

\nsection{Lecture 1. \,Wess-Zumino amplitudes and bundle gerbes}
\label{sec:lect1}
\subsection{Introduction}
\label{subsec:intro}

\noindent These lectures are devoted to the application 
of techniques related to gerbes to the construction of torsion invariants
of low-dimensional topological insulators. In recent times,
the subject of topological insulators has been an important point of junction 
between condensed matter theory and mathematics. The interaction started 
from the realizations of the role that the $1^{\rm st}$ Chern number plays 
in the integer quantum Hall effect \cite{TKNN,ASS} and the relations 
of the later to the index theorem. It gained a new momentum with the 
introduction of $K$-theoretic invariants to classify time-reversal topological
insulators \cite{KM,Kitaev}. In these lectures I shall present
a geometric picture of the simplest of those $K$-theoretic invariants,
the 2- and 3-dimensional Kane-Mele $\mathbb Z_2$-valued index
and its generalization to the so called Floquet topological insulators 
that we proposed in \cite{CDFG,CDFGT}. The geometric picture is centered
on the concept of Wess-Zumino amplitudes and their refinements and on
the underlying geometry of bundle gerbes and equivariant structures on them. 
These concepts will be the topic of the first lecture preparing the ground 
for the second one where I shall describe how they are applied to construct 
indices classifying time-reversal invariant topological insulators.

\subsection{$2d$ Wess-Zumino amplitudes and their square root}
\label{subsec:WZamplis}
        
\noindent Let us start by recalling what are the Wess-Zumino amplitudes
\cite{WZ,Wit}. Let $\,M\,$ be a manifold and $\,H\,$ a closed 
real $(k+1)$-form on $M$ whose periods, i.e. integrals over singular 
$(k+1)$-cycles, are in $2\pi\mathbb Z$. Mathematically, the Wess-Zumino (WZ) 
amplitude is a Cheeger-Simons differential character, i.e. a homomorphism 
\qq
Z_k(M)\,\longrightarrow\,\mathbb R/(2\pi\mathbb Z)
\qqq
from the group of singular $k$-cycles in $\,M\,$ to $\,U(1)\,$ that takes on 
boundaries $\,\partial  c_{k+1}\,$ the values 
$\,\int_{c_{k+1}}\hspace{-0.3cm}H\ \,{\rm mod}\,\,2\pi\mathbb Z\,$ 
\cite{CheegSim}. The differential characters (with arbitrary 
$\,(k+1)$-forms $\,H$) \,form an Abelian group $\,\hat H^{k+1}(M)$.
\,We shall consider here the case with $k=2$ where $\,H\,$ is 
a 3-form and we shall use physicist's notation writing 
\qq
\ee^{\ii S_{\WZ}(\phi)}\in U(1)
\qqq
for the values of the differential character on a singular
2-cycle $\,c_\phi\,$ determined by a smooth map $\,\phi:\Sigma\rightarrow M\,$
from a closed oriented 2-surface $\,\Sigma\,$ to $\,M$. It follows
then from the definition that if there exist an oriented $3d$-manifold
$\,\widetilde\Sigma\,$ such that $\,\partial\widetilde\Sigma=\Sigma\,$
and a smooth map $\,\widetilde\phi:\widetilde\Sigma\rightarrow M\,$
extending $\,\phi$, i.e. such that 
$\,\widetilde\phi|_{\partial\widetilde\Sigma}=\phi$, then
\qq
\ee^{\ii S_{\WZ}(\phi)}=\exp\Big[\ii\int_{\widetilde\Sigma}
\widetilde\phi^*H\Big].
\label{WZampli}
\qqq
Differential characters corresponding to a given 3-form $\,H\,$
differ by elements $\,\chi\in Hom(H_2(M),U(1))\cong H^2(M,U(1))$, or more 
exactly, by multiplication by $\,\chi([c_\phi])$, where $\,[c_\phi]\,$ denotes 
the homology class of the 2-cycle $\,c_\phi$. 
\vskip 0.2cm

We shall be interested in the case where $M=U(N)$ and $H$ is the closed 
bi-invariant 3-form,  
\qq
H=\frac{_1}{^{12\pi}}\,\tr(g^{-1}dg)^{\wedge 3}\,,
\label{H}
\qqq
that is normalized so that its set of periods is $\,2\pi\mathbb Z$. Since 
$H_2(U(N))=0$, the WZ amplitudes are uniquely fixed by the rule (\ref{WZampli})
as every $\,\phi\,$ extends to $\,\widetilde\phi\,$ in this case.
\vskip 0.2cm

The main idea pursued in this lectures is that the presence of time reversal 
symmetry imposes the consideration of square roots of WZ amplitudes. 
In quantum mechanics the time reversal is realized by an anti-unitary
operator in the space of states. In particular, in the space of states
$\mathbb C^N\cong\mathbb C^{\otimes_s2j}$ with $N=2j+1$ carrying
the representation of spin $j=0,\frac{1}{2},1,\dots$, \,the time reversal
is realized by the anti-unitary operator $\theta=\ee^{\pi\ii S_y}C$, where
$C$ is the complex conjugation and $S_y$ is the $y$-component of the
spin operator $\vec{S}$. In this case $\theta^2=(-1)^{2j}I$. We shall 
be mostly interested in the case when $\theta^2=-I$ corresponding to
half-integer spins, e.g. to $j=\frac{1}{2}$ as for electrons.
The adjoint action of $\theta$ induces an involution 
$\Theta:U(N)\rightarrow U(N)$,
\qq
\Theta(V)=\theta V\theta^{-1},
\qqq
which preserves the bi-invariant 3-form $H$ on $U(N)$. Suppose now that 
we equip the closed oriented surface $\Sigma$ with a nontrivial 
orientation-preserving involution $\vartheta$ with discrete non-empty 
set of fixed points. The typical example will be the action induced on 
the torus $\,\mathbb T^2=\mathbb R^2/(2\pi\mathbb Z)^2\,$ by the map 
$\,k\mapsto -k$ on $\mathbb R^2$. Other examples may be given by
the map $(z,y)\mapsto(z,-y)$ on the hyperelliptic curve $y^2=p(z)$,
where $\,p\,$ is a polynomial. Let $\phi:\Sigma\mapsto U(N)$ be an 
equivariant map, i.e. such that
\qq
\phi\circ\vartheta =\Theta\circ\phi\,.
\label{thetaphi}
\qqq
Suppose that there exists an oriented $3d$ manifold $\widetilde\Sigma$ 
equipped with an orientation-preserving involution $\widetilde\vartheta$ 
such that $\,\partial\widetilde\Sigma=\Sigma\,$ and 
$\,\widetilde\vartheta|_{\partial\widetilde\Sigma}=\vartheta\,$ and an extension 
$\,\widetilde\phi:\widetilde\Sigma\rightarrow U(N)\,$ of $\,\phi\,$ such that
\qq
\widetilde\phi\circ\widetilde\vartheta=\Theta\circ\widetilde\phi\,.
\label{thetatildephi}
\qqq
Let us set
\qq
\sqrt{\ee^{\ii S_\WZ(\phi)}}\,=\,
\exp\Big[\frac{_\ii}{^2}\int_{\widetilde\Sigma}\widetilde\phi^*H\Big].
\label{sqrtWZ}
\qqq
Does this provide a correct definition of the square root of the WZ amplitude
of equivariant maps $\phi:\Sigma\rightarrow U(N)$?
\vskip 0.3cm

\noindent{\bf Proposition 1.} \ Assume that $\,\theta^2=-I$. Let 
$\,\Sigma=\mathbb T^2$, $\,\vartheta k=-k$, \,and let $\,\phi:
\Sigma\rightarrow U(N)\,$ satisfy (\ref{thetaphi}). Applying at most 
an $\,SL(2,\mathbb Z)\,$ change of 
variables on $\,\Sigma$, we may assume that $\,\det(\phi)\,$ does not wind 
around the first circle of $\,\mathbb T^2=S^1\times S^1$. Let 
$\,\tilde\Sigma=D\times S^1$, where $\,D\,$ is a unit disk in $\,\mathbb C$, 
\,with $\,\widetilde\vartheta(z,\lambda)=(\bar z,\bar\lambda)$. Then there 
exists an extension of $\,\widetilde\phi:\widetilde\Sigma\rightarrow U(N)\,$ 
of $\,\phi\,$ satisfying (\ref{thetatildephi}) and the right hand side of 
(\ref{sqrtWZ}) does not depend on its choice so that $\sqrt{\ee^{\ii S_\WZ(\phi)}}$ 
given by (\ref{sqrtWZ}) is well defined. 
\vskip 0.3cm

\noindent{\bf Remarks.} \ 1. The essence of the last statement is that 
the imposition of condition (\ref{thetatildephi}) on $\,\widetilde\phi\,$
makes $\,\int_{\widetilde\Sigma}\widetilde\phi^*H\,$ well defined modulo 
$\,4\pi\,$ rather than only modulo $\,2\pi\,$ which would be the case 
without that restriction. The assumption $\,\theta^2=-I\,$ is essential 
for both assertions of Proposition 1.

\noindent 2. The proof of that proposition is rather laborious. It is 
a simple extension of the one given in \cite{CDFGT} where it was assumed
that $\,\det(\phi)\,$ has no windings.
\vskip 0.3cm

The applications of the above construction will be discussed in
Lecture 2. Let us only mention here that, when applied to the 
map $\,\phi:\mathbb T^2\rightarrow U(N)$, $\,\phi(k)=I-2P(k)$, \,where 
$\,P(k)=\theta P(-k)\theta^{-1}\,$ is the family of projectors on 
the valence band states of a time-reversal invariant $2d$ insulator, 
\,it gives
\qq
\sqrt{\ee^{\ii S_\WZ(\phi)}}=(-1)^{\KM},
\label{WZKM}
\qqq
where $\KM\in\mathbb Z_2$ is the Kane-Mele invariant \cite{KM,FK}
of such insulators.
\vskip 0.2cm

Although the nonlocal expressions (\ref{WZampli}) for 
$\,\ee^{\ii S_\WZ(\phi)}\,$ and (\ref{sqrtWZ}) 
for $\sqrt{\ee^{\ii S_\WZ(\phi)}}$ in terms of $3d$ integrals 
define well those quantities in the cases described above, we shall 
also need local expressions for them in terms of $2d$ integrals with
corrections. Such expressions may be conveniently described using 
the holonomy of bundle gerbes and its appropriate refinement. The 
local expressions are essential for the construction of a $3d$ index
that will be presented in Sec.\,\ref{subsec:3d_ind} below. Such
index will be our main tool for building topological invariants 
of insulators.

\subsection{Bundle gerbes with connection}
\label{subsec:bdlgerbes}

\noindent Bundle gerbes were introduced by Murray in \cite{Murray} as
a geometric example of more abstract gerbes considered in
\cite{Giraud} and \cite{Brylinski}. I shall follow closely 
the original definition of \cite{Murray}.
\vskip 0.2cm
 
Let $\,\pi:Y\mapsto M\,$ be a surjective submersion and 
\qq
Y^{[n]}\,=\,Y\times_MY\times_M\cdots\,\times_MY
\,=\,\big\{(y_1,\dots y_n)\in Y^{n}\,\big|\,\pi(y_1)=\cdots=\pi(y_n)\big\}.
\qqq 
By $\,p_{i_1\cdots i_m}:Y^{[n]}\rightarrow Y^{[m]}\,$ we shall denote the maps 
$\,(y_1,\dots y_n)\mapsto(y_{i_1},\dots y_{i_m})$.
\vskip 0.3cm

\noindent{\bf Definition} \cite{Murray}{\bf.} \ \,A bundle gerbe with 
connection $\,\CG\,$ over $\,M\,$ is a 
quadruple $\,(Y,B,\CL,t)$, \,where $\,\pi:Y\rightarrow M\,$ is a surjective 
submersion, $\,B\,$ is a 2-form
on $\,Y\,$ (called curving), $\,\CL\,$ is a Hermitian line bundle with unitary
connection\footnote{All line bundles considered here are assumed to be 
equipped with such structures and their isomorphisms to preserve them, 
unless stated otherwise.} over $Y^{[2]}$ with (real) curvature 2-form 
$\,p_1^*B-p_0^*B$, $\,t\,$ is a line-bundle isomorphism over $\,Y^{[3]}\,$
\qq
t:p_{12}^*\CL\otimes p_{23}^*\CL\longrightarrow p_{13}^*\CL
\qqq
acting fiberwise\footnote{We denote by $\,\CL_{y_1,y_2}\,$ the fiber
of $\,\CL\,$ over $\,(y_1,y_2)\in Y^{[2]}$.} as
\qq
\CL_{y_1,y_2}\otimes\CL_{y_2,y_3}\,\mathop{\longrightarrow}\limits^t\,
\CL_{y_1,y_3}
\label{tfibrewise}
\qqq
for $(y_1,y_2,y_3)\in Y^{[3]}$ and defining an (associative) groupoid 
multiplication on $\,\CL\,\substack{\rightarrow\\\rightarrow}\,Y$.
In particular, $\,t\,$ provides canonical isomorphisms $\,\CL_{y,y}
\cong\mathbb C\,$ and $\,\CL_{y_1,y_2}^{-1}\cong\CL_{y_2,y_1}$, \,where
$\,\CL_{y_1,y_2}^{-1}\,$ denotes the line dual to $\,\CL_{y_1,y_2}$.
\vskip 0.3cm

\noindent The condition on the curving
2-form implies that $\,p_1^*dB=p_0^*dB\,$ so that $\,dB=\pi^*H\,$ for some
closed 3-form $\,H\,$ on $\,M\,$ called the curvature of gerbe $\,\CG$.
\vskip 0.2cm

Gerbes over $\,M\,$ form a 2-category with objects, 1-isomorphisms between
objects and 2-isomorphisms between 1-isomorphisms \cite{Stev}. We  shall
only need 1-isomorphisms $\,\eta:\CG\rightarrow\CG'\,$
between gerbes $\,\CG=(Y,B,\CL,t)\,$ and $\,\CG'=(Y,B',\CL',t')\,$ with 
the same $Y$. They may be given by a line bundle $\,\CN\,$ over $\,Y\,$ 
with curvature $\,B'-B$, and by an isomorphism $\,\nu\,$ of line bundles 
over $\,Y^{[2]}$  
\qq
\nu:\CL\otimes p_2^*\CN\,\longrightarrow\,p_1^*\CN\otimes\CL'
\label{n}
\qqq
intertwining the groupoid multiplications $\,t\,$ and $\,t'$. 
\,In particular, $\,\nu\,$ establishes an isomorphism
\qq
\CL_{y_1,y_2}\otimes\CL'^{-1}_{y_1,y_2}\cong\,\CN_{y_1}\otimes\CN_{y_2}^{-1}.
\label{nufw}
\qqq
1-isomorphic 
gerbes have the same curvature. A 2-isomorphism $\,\mu:\eta^1
\Rightarrow\eta^2\,$ between 1-isomorphisms $\,\eta^\alpha:\CG\rightarrow\CG'$, 
$\alpha=1,2$, $\,\eta^\alpha=(\CN^\alpha,\nu^\alpha)\,$ 
is an isomorphism $\,m:\CN^1\rightarrow\CN^2\,$ of line bundles over $Y$ that 
makes commutative the diagram
\qq
&&\qquad\ \ \CL\otimes p_2^*\CN^1\ \mathop{\longrightarrow}^{\nu^1}\ 
p_1^*\CN^1\otimes\CL'\cr\cr
&&\Id_\CL\otimes p_2^*m\Big\downarrow\hspace{2.3cm}\Big\downarrow p_1^*m\otimes
\Id_{\CL'}\cr\cr
&&\qquad\ \ \CL\otimes p_2^*\CN^2\ \mathop{\longrightarrow}^{\nu^2}
\ p_1^*\CN^2\otimes\CL'
\qqq 
\vskip 0.2cm

Gerbes $\,\CG$, \,1-isomorphisms $\,\eta\,$ between them and 2-isomorphisms 
$\,\mu\,$ can be multiplied and composed in natural ways and pulled back
by maps $\,T:M'\rightarrow M$. \,The 1-isomorphism classes of
gerbes over $\,M\,$ form an Abelian group $\,\mathbb G(M)$.
\vskip 0.4cm

\noindent{\bf Remark.} \ If $\,\eta:\CG\rightarrow\CG$, $\,\eta=(\CN,\nu)$, 
\,is a 1-isomorphism of the same gerbe over $\,M\,$ then line bundle
isomorphism $\,\nu\,$ allows to view $\,\CN\,$ as a pullback of a flat bundle 
$\,N\,$ over $\,M$. The identity 1-isomorphism $\,\Id_\CG$ with trivial bundle 
$\,\CN\,$ and identity isomorphism $\,\nu\,$ is an example of such a 
1-isomorphism. There exists a 2-isomorphism $\,\mu:\eta\Rightarrow\Id_\CG\,$ 
if and only if the flat line bundle $\,N\,$ corresponding to $\,\CN\,$ is 
trivializable, with the trivialization defining the isomorphism $\,m\,$ 
corresponding to $\,\mu$.

\subsection{Bundle gerbe holonomy}
\label{subsec:gerbe_hol}

\noindent Group $\mathbb G(\Sigma)$ of 1-isomorphism classes of gerbes
over a closed oriented surface $\Sigma$ is isomorphic to $U(1)$. If
$\CG=(Y,B,\CL,t)$ is a gerbe over $M$ and $\phi:\Sigma\rightarrow M$ then 
the phase in $U(1)$ associated to the 1-isomorphism class of gerbe $\phi^*\CG$ 
over $\Sigma$ is called the holonomy of $\CG$ along $\phi$ and is
denoted $Hol_\CG(\phi)$. We shall need an explicit representation
of such a phase. 
\vskip 0.1cm

\begin{figure}[!t]
\vskip 0.2cm
\begin{center}
\leavevmode
 \includegraphics[width=10cm,height=4.5cm]{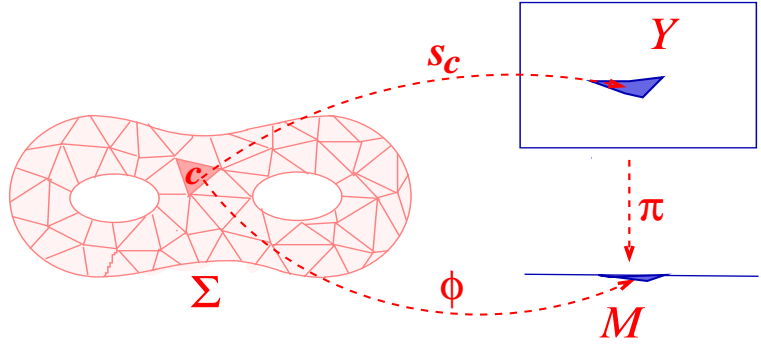}
\vskip -0.1cm
\caption{Triangulation of $\,\Sigma\,$ for gerbe holonomy calculation}
\label{fig:triang0}
\end{center}
\vskip -0.2cm
\end{figure}

\begin{figure}[!h]
\vskip 0.2cm
\begin{center}
\leavevmode
 \includegraphics[width=5.4cm,height=3.4cm]{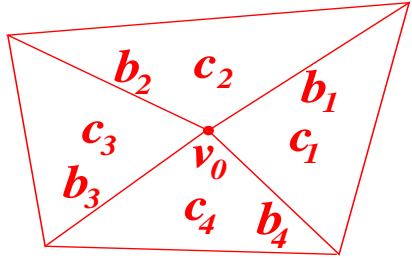}
\vskip -0.1cm
\caption{Triangulation of $\,\Sigma\,$ around vertex $\,v_0$}
\label{fig:aroundv0}
\end{center}
\vskip -0.7cm
\end{figure}

Let us choose a triangulation of $\Sigma$ composed of triangles
$\,c$, \,edges $\,b\,$ and vertices $\,v$. \,We suppose that it is
sufficiently fine so that there exist maps 
$s_c:c\rightarrow Y$ such that  
\qq 
\pi\circ s_c=\phi|_c\,,
\label{lsc}
\qqq
see Fig.\,\ref{fig:triang0}. For each edge we shall also choose
a a map $s_b:b\mapsto Y$ such that $\pi\circ s_b=\phi|_b$ and for
each vertex $v$ a point $s_v\in Y$ such that $\pi(s_v)=\phi(v)$.
Then the holonomy of $\CG$ along $\phi$ may be given by the expression
\cite{GR}
\qq
Hol_\CG(\phi)\,=\,\ee^{i\sum\limits_c\int_cs_c^*B}
\mathop{\otimes}\limits_{b\subset c}hol_\CL(s_c|_b,s_b)\,, 
\label{HolCG}
\qqq
where we use a slightly abusive notation in which $hol_\CL(\ell)$
stands for the parallel transport of in line bundle $\CL$ along
curve $\ell$ in $Y^{[2]}$ which is a linear map from the fiber 
of $\CL$ over the initial point of $\ell$ to the one over the final 
point. It follows that the expression on the right hand side of
(\ref{HolCG}) is an element of the line
\qq
\mathop{\otimes}\limits_{v\in b\subset c}\CL_{s_c(v),s_b(v)}^{\pm1},
\label{line}
\qqq
where the minus power (the dual line) is chosen if $v$ has a negative
orientation, i.e. is the initial point of the edge $b$ with orientation 
inherited from $c$. The groupoid structure on $\,\CL\,$ defined by the 
isomorphism $\,t\,$ of (\ref{tisom}) makes the line (\ref{line}) canonically 
isomorphic to $\mathbb C$. Indeed, for a fixed vertex $v_0$ as
in Fig.\,\ref{fig:aroundv0},

\qq
\mathop{\otimes}\limits_{v_0\in b\subset c}\CL_{s_c(v),s_b(v)}^{\pm1}
&\cong&\CL_{s_{c_1}(v_0),s_{b_1}(v_0)}\otimes\CL_{s_{b_1}(v_0),s_{c_2}(v_0)}
\otimes\CL_{s_{c_2}(v_0),s_{b_2}(v_0)}\otimes\CL_{s_{b_2}(v_0),s_{c_3}(v_0)}\cr
&&\hspace{-1.4cm}\otimes\,\CL_{s_{c_3}(v_0),s_{b_3}(v_0)}\otimes\CL_{s_{b_3}(v_0),
s_{c_4}(v_0)}\otimes\CL_{s_{c_4}(v_0),s_{b_4}(v_0)}\otimes\CL_{s_{b_1}(v_0),s_{c_1}(v_0)}
\ \cong\ \mathbb C\,.\quad
\qqq
A cyclic permutation of terms does not change the isomorphism with $\mathbb C$
because of the associativity of the groupoid multiplication in $\CL$. 
Hence, the right hand side of (\ref{HolCG}) may be canonically viewed
as a complex number that, in fact, belongs to $U(1)$. The holonomy of
gerbe with curvature $H$ is a differential character that we shall use 
to define the WZ amplitude corresponding to the closed 3-form $H$
setting
\qq
\ee^{\ii S_\WZ(\phi)}\,=\,Hol_\CG(\phi)\,.
\label{WZHol}
\qqq    
\vskip 0.1cm   

For the later use, let us notice that if we used the same formula
(\ref{HolCG}) for a surface $\Sigma$ with boundary $\partial\Sigma\cong S^1$
then we would obtain
\qq
Hol_\CG(\phi)\,\in\,\mathop{\otimes}_{v\in b\subset\partial\Sigma}
\CL_{s_v,s_b(v)}^{\pm1}\cong\ \CL^{\CG}_{\phi|_{\partial\Sigma}}\,,
\label{Holbd}
\qqq 
where $\,\CL^\CG\,$ is the transgression line bundle over loop space 
$\,LM\,$ canonically induced by gerbe $\CG$ over $M$ \cite{GR}.

\subsection{Gerbes equivariant under involution}
\label{subsec:equivariant_gerbes}

\noindent Equivariance of gerbes has been studied by several authors,
see \cite{Gomi,Gomi1,SSW,GSW,BenBassat,MRSV}. We shall discuss here
a simple version of such equivariance under an action of $\,\mathbb Z_2\,$ 
group.
\vskip 0.2cm

\noindent Let $\Theta:M\rightarrow M$ be an involution and
let $\CG=(Y,B,\CL,t)$ be a bundle gerbe over $M$ with curvature $H$. 
We shall assume that $\Theta$ may be lifted to an involution 
$\Theta_Y:Y\rightarrow Y$ covering $\,\Theta$, \,i.e. such that 
the diagram
\qq
&&\ \ Y\ \,\mathop{\longrightarrow}\limits^{\Theta_Y}\,\ Y\cr
&&\pi\downarrow\hspace{1.1cm}\downarrow\pi\cr
&&\ \ M\ \mathop{\longrightarrow}\limits^{\Theta}\ M
\qqq
is commutative. We would like to compare gerbe $\CG$ to its pullback
$\Theta^*\CG$ that can be realized as the quadruple 
$\,(Y,\Theta_Y^*B,(\Theta^{[2]}_Y)^*\CL,(\Theta^{[3]}_Y)^*t)$. 
\vskip 0.4cm

\noindent{\bf Definition.} \ $\Theta$-equivariant structure
on $\CG$ is a pair $(\eta,\mu)$ where $\eta=(\CN,\nu)$ is a
1-isomorphism $\,\eta:\CG\rightarrow\Theta^*\CG\,$
and $\,\mu:\Theta^*\eta\circ\eta\Rightarrow\Id_\CG\,$ is a 
2-isomorphism between 1-isomorphisms of $\,\CG$. 
\,Besides, as 2-isomorphisms between 1-isomorphisms 
$\eta\circ\Theta^*\eta\circ\eta:\CG\rightarrow\Theta^*\CG$ 
and $\eta:\CG\rightarrow\Theta^*\CG$,
\qq
\Theta^*\mu\circ\Id_\eta\,=\,\Id_\eta\circ\mu\,.
\label{mu2}
\qqq 
\vskip 0.4cm

\noindent{\bf Remarks.} \ 1. The existence of a 1-isomorphism $\,\eta:\CG\rightarrow
\Theta^*\CG\,$ implies that $\,H=\Theta^*H$. 

\noindent 2. The composition $\,\Theta^*\eta\circ\eta=(\CQ,\rho)$
with $\CQ=\Theta_Y^*\CN\otimes\CN$ is a 1-isomorphism of gerbe $\CG$ and,
consequently, $\CQ=\pi^*Q$, where $Q$ is a flat line bundle over $M$.
Line bundle $Q$ comes with an involutive isomorphism $\Theta_Q$ 
switching the tensor factors that covers $\Theta$. The 2-isomorphism $\mu$ 
is given by a trivialization of $Q$ defined by a flat normalized section 
$S:M\rightarrow Q$, $|S|=1$. Relation (\ref{mu2}) translates to the condition
\qq
\Theta_Q\circ S=S\circ\Theta\,.
\label{S2}
\qqq

\subsection{Square root of gerbe holonomy}
\label{subsec:sqrt_hol}

\noindent Under special conditions that will be specified below, a 
$\Theta$-equivariant structure $(\eta,\mu)$ on a gerbe $\CG=(Y,B,\CL,t)$ 
over $M$ permits to determine a square root of the holonomy $Hol_\CG(\phi)$ 
of maps $\phi:\Sigma\rightarrow M$ that satisfy the equivariance condition
(\ref{thetaphi}) for an orientation-preserving involution $\vartheta:\Sigma
\rightarrow\Sigma$ with discrete fixed points.  
\vskip 0.1cm

\begin{figure}[b]
\begin{center}
\vskip -0.1cm
\leavevmode
\hspace{1.2cm}
{%
      \begin{minipage}{0.43\textwidth}
        \includegraphics[width=4.5cm,height=3.6cm]{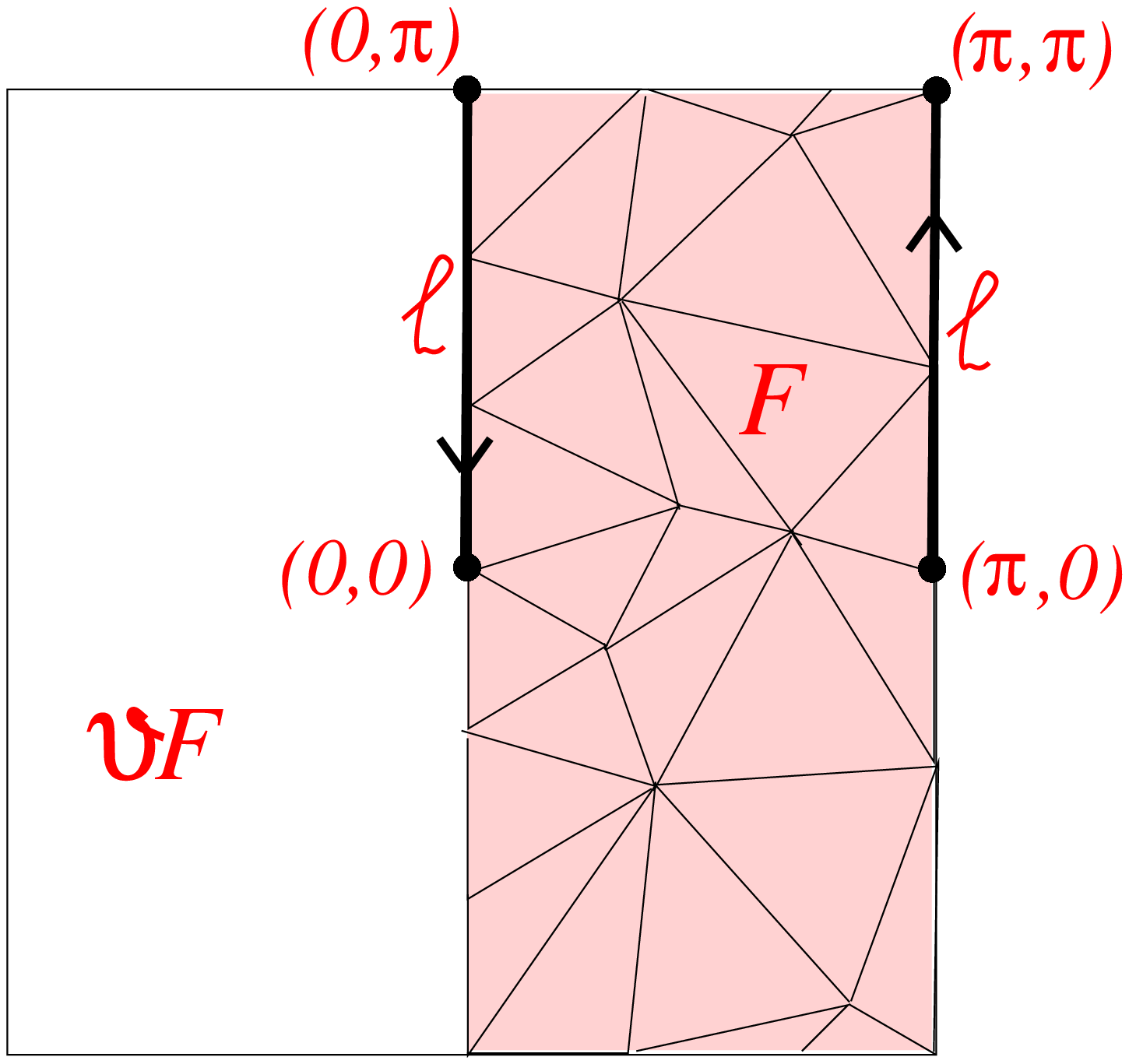}\\
        \vspace{-0.8cm} \strut
        \end{minipage}}
    \hspace*{-0.8cm}
{%
      \begin{minipage}{0.43\textwidth}
        \includegraphics[width=4.4cm,height=3.6cm]{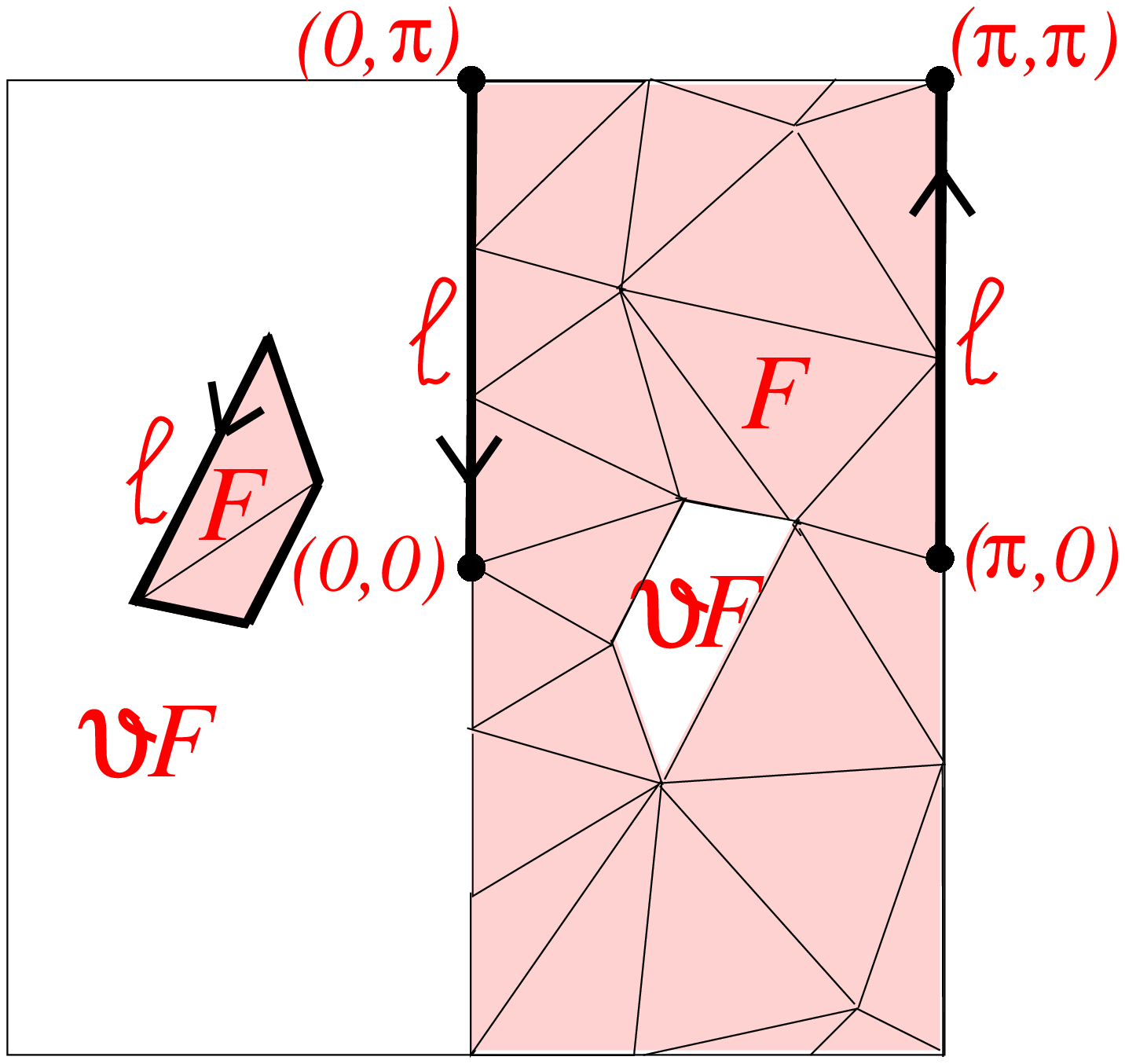}\\
      \vspace{-0.8cm} \strut
        \end{minipage}}
      \hspace*{15pt}
    \hspace{14pt}
\end{center}
\caption{Two examples of fundamental domain $\,F\,$ for $\,\Sigma
=\mathbb T^2$}
\label{fig:triangul}
\end{figure}

Let us choose a fundamental domain in $\,\Sigma\,$ for $\vartheta$ whose
closure $\,F\,$ is a (piecewise) smooth submanifold with boundary of 
$\,\Sigma$, \,see Fig.\,\ref{fig:triangul} for examples of possible choices 
of $\,F\,$ for $\Sigma=\mathbb T^2$.

\noindent We shall triangulate $\,F\subset\Sigma\,$ in a way that is 
$\,\vartheta$-symmetric when restricted to $\,\partial F\,$ ($\vartheta\,$
preserves $\,\partial F\,$ reversing its orientation). Assuming that
the triangulation of $\,F\,$ is sufficiently fine so that the 
maps $\,s_c\,$ satisfying (\ref{lsc}) exist, we shall first
consider
\qq
Hol_\CG(\phi|_F)\,\in\,\mathop{\otimes}_{v\in b\subset\partial F}
\CL_{s_v,s_b(v)}^{\pm1}
\qqq 
according to (\ref{Holbd}). Let us now choose a fundamental domain
for $\,\vartheta\,$ acting in $\,\partial F\,$ and let $\,\ell\,$ \
be its closure. Connected components of $\,\ell\,$ are either
curves beginning and ending at fixed points of $\,\vartheta\,$
or closed loops, \,see Fig.\,\ref{fig:triangul}. The maps
$\,s_b:b\rightarrow Y\,$ such that $\,\pi\circ s_b=\phi|_b\,$ 
may be chosen so that for $\,b\subset\ell$,
\qq
s_{\vartheta(b)}\circ\vartheta=\Theta_Y\circ s_b\,.
\label{sbsb}
\qqq   
Similarly, we may choose $\,s_v\in Y\,$ for vertices $\,v\,$ in the
interior of $\,\ell\,$ so that
\qq
s_{\vartheta(v)}=\Theta_Y(s_v)\,.
\qqq 
In this case,
\qq
\mathop{\otimes}_{v\in b\subset \partial F}\CL_{s_v,s_b(v)}^{\pm1}&\cong&
\Big(\mathop{\otimes}\limits_{v\in b\subset\ell}
\big(\CL_{s_v,s_b(v)}\otimes\CL_{\Theta_Y(s_v),\Theta_Y(s_b(v))}^{-1}\big)^{\pm1}\Big)
\otimes\Big(\mathop{\otimes}\limits_{v\in\partial\ell}\CL_{\Theta_Y(s_v),s_v}^{\pm 1}\Big)\cr
&\cong&\Big(\mathop{\otimes}\limits_{v\in b\subset\ell}
\big(\CN_{s_v}\otimes\CN_{s_b(v)}^{-1}\big)^{\pm1}\Big)\otimes
\Big(\mathop{\otimes}\limits_{v\in\partial\ell}\CL_{\Theta_Y(s_v),s_v}^{\pm 1}\Big)\cr
&\cong&\Big(\mathop{\otimes}\limits_{v\in b\subset\ell}
\CN_{s_b(v)}^{\mp1}\Big)\otimes
\Big(\mathop{\otimes}\limits_{v\in\partial\ell}
\big(\CL_{\Theta_Y(s_v),s_v}\otimes\CN_{s_v}\big)^{\pm 1}\Big),
\label{1lineisom}
\qqq 
where the second line is obtain using the line bundle isomorphism 
$\,\nu\,$ of the 1-isomorphism $\,\eta:\CG\rightarrow\Theta^*\CG$,
\,see relation (\ref{nufw}). 
\vskip 0.1cm

In the next step, let us set, abusing again notations, 
\qq
hol_\CN(\phi|_\ell)\,=\,\mathop{\otimes}\limits_{b\subset\ell}hol_\CN(s_b)
\ \in\,\mathop{\otimes}\limits_{v\in b\subset\ell}
\CN_{s_b(v)}^{\pm1}\,,
\qqq
where $\,hol_\CN(s_b)\,$ stands for the parallel transport in line
bundle $\,\CN\,$ along $\,s_b$. \,We infer that
\qq
Hol_\CG(\phi|_F)\otimes hol_{\CN}(\phi|_\ell)\,\in\,
\mathop{\otimes}\limits_{v\in\partial\ell}
\big(\CL_{\Theta_Y(s_v),s_v}\otimes\CN_{s_v}\big)^{\pm 1}
\label{Holhol}
\qqq
The latter expression will eventually give the value of the square root
of $\,Hol_\CG(\phi)\,$ after the identification of lines 
$\,\CL_{\Theta_Y(s_v),s_v}\otimes\CN_{s_v}\,$
with $\,\mathbb C\,$ that we shall discuss now. 
\vskip 0.1cm

Let us first note that $\,\phi(\partial\ell)\subset M'$, \,where
\qq
M'\,=\,\big\{x\in M\,\big|\,\Theta(x)=x\big\} 
\qqq
is the fixed point set of $\,\Theta\,$ that, for simplicity, we shall assume 
to be a submanifold of $\,M$. \,We shall denote by $\,Y'\subset Y\,$ the
preimage by $\,\pi\,$ of $\,M'$. \,Note that $\,s_v\in Y'\,$ for 
$\,v\in\partial\ell$. \,Let $\,r:Y'\rightarrow Y'^{[2]}\,$
be defined by $\,r(y')=(\Theta_Y(y'),y')$. \,Consider the line bundle 
\qq
\CN'\,=\,r^*\CL\otimes \CN|_{Y'}
\label{CN'}
\qqq
over $\,Y'\,$ with fibers
\qq
\CN_{y'}\,=\,\CL_{\Theta_Y(y'),y'}\otimes\CN_{y'}\,.
\qqq
Note that $\,\CN'\,$ is flat. Besides, it has a natural structure of a pullback 
of a flat line bundle $\,N'$ over $\,M'\,$ given by the identification of its fibers
over points $\,y'_1\,$ and $\,y'_2\,$ with $\,\pi(y'_1)=\pi(y'_2)$:
\qq
&&\CN'_{y'_1}=\CL_{\Theta_Y(y'_1),y'_1}\otimes\CN_{y'_1}\,\mathop{\longrightarrow}
\limits^{t^{-1}\otimes\Id_\CN}\,\CL_{\Theta_Y(y'_1),y'_2}\otimes\CL_{y_2',y'_1}
\otimes\CN_{y'_1}\cr
&&\mathop{\longrightarrow}\limits^{\Id_\CL\otimes\nu}\,
\CL_{\Theta_Y(y'_1),y_2}\otimes\CN_{y'_2}\otimes\CL_{\Theta_Y(y'_2),\Theta_Y(y'_1)}
\,\cong\,\CL_{\Theta_Y(y'_2),\Theta_Y(y'_1)}\otimes\CL_{\Theta_Y(y'_1),y_2}\otimes
\CN_{y'_2}\cr
&&\mathop{\longrightarrow}\limits^{t\otimes\Id_\CN}\,
\CL_{\Theta_Y(y'_2),y_2}\otimes\CN_{y'_2}=\CN'_{y'_2}\,.
\label{CN'N'}
\qqq
$\,\CN'\,$ together with this structure represents the 1-isomorphism
$\,\eta|_{M'}:\CG'\rightarrow\CG'$, \,where $\,\CG'=\CG|_{M'}$.
In particular, the above identification defines a line bundle
isomorphism $\,\nu':\CN'\rightarrow\Theta_Y^*\CN'$. \,Now
\qq
\CN'\otimes\CN'\,\cong\,\Theta_Y^*\CN'\otimes\CN'\,\cong\,
(\Theta_Y^*\CN\otimes\CN)|_{M'}\,=\,\CQ|_{M'}\,\equiv\,\CQ'\,,
\label{CN'2}  
\qqq
where $\,\CQ\,$ is the flat line bundle over $\,Y\,$ corresponding to the 
1-isomorphism $\,\Theta^*\eta\circ\eta:\CG\rightarrow\CG$.
\,Recall that $\,\CQ\cong\pi^*Q\,$ where $\,Q\,$ is a flat line bundle
over $\,M\,$ provided with a flat section $\,S\,$ as a part of the
$\,\Theta$-equivariant structure on gerbe $\,\CG$. \,Thus section 
$\,S'=S|_{M'}\,$ may be viewed as providing a trivialization of the flat 
line bundle $\,N'\otimes N'$. If $\,M'\,$ is simply connected 
then there also exists section $\,\sqrt{S'}\,$ trivializing flat line 
$\,N'\,$ such that $\,\sqrt{S'}\otimes\sqrt{S'}=S'\,$ (all flat line
bundles on simply connected manifolds are trivializable). Besides, 
$\,\sqrt{S'}\,$ is determined modulo a locally constant function
taking values in $\,\{\pm1\}$. \,In particular, if $\,M'\,$ is simply
connected, the trivialization $\,\sqrt{S'}\,$ of $\,N'\,$ is defined 
up to a global sign. \,From (\ref{Holhol}) it follows that if
$\,M'\,$ is connected and simply connected then
\qq
Hol_\CG(\phi|_F)\otimes hol_{\CN}(\phi|_\ell)\,\in\,
\mathop{\otimes}\limits_{v\in\partial\ell}N'_{\phi(v)}\,\cong\,\mathbb C
\label{sqrtHol}
\qqq
and the last isomorphism using section $\,\sqrt{S'}\,$ of $\,N'\,$
is independent of the choice of that section (the global sign change
cancels between the terms coming from the two ends of connected components 
of $\,\ell$).  
\vskip 0.4cm

\noindent{\bf Proposition 2.} \ For $\,M'\,$ connected and simply connected,
the latter isomorphism associates to $\,Hol_\CG(\phi|_F)\otimes 
hol_{\CN}(\phi|_\ell)\,$ a phase in $\,U(1)\,$ that does not depend on 
the choices of $\,F$, $\,\ell$, \,the triangulation
of $\,F$, the lifts $\,s_c$, $\,s_b\,$ and $\,s_v\,$ and of the sign
of $\,\sqrt{S'}$.  
\vskip 0.4cm

\noindent {\bf Remark.} \ The independence of the phase assigned
to $\,Hol_\CG(\phi|_F)\otimes 
hol_{\CN}(\phi|_\ell)\,$ on the choices of $\,s_c$, $\,s_b\,$ and $\,s_v\,$
is proven by a direct check, that on the triangulation by passing to
a finer one with respect to two arbitrary triangulations, on the choice of $\,\ell\,$
by replacing one of its connected components by its $\vartheta$ image
and on $\,F\,$ by replacing one of its triangles by its $\vartheta$ image.   
Multiplying the expressions corresponding to fundamental domains
$\,F\,$ and $\,\vartheta F\,$ one restores $\,Hol_{\CG}(\phi)\,$
proving that the phase in question squares to $\,Hol_{\CG}(\phi)$.
\vskip 0.4cm

\noindent{\bf Definition.} \ Under the assumptions of Proposition 2,
\,we define $\,\sqrt{Hol_\CG(\phi)}\,$ by the relation 
\qq
Hol_\CG(\phi|_F)\otimes hol_{\CN}(\phi|_\ell)\,=\,
\sqrt{Hol_\CG(\phi)}\,\mathop{\otimes}\limits_{v\in\partial\ell}
\sqrt{S'}(\phi(v))^{\pm1}\,.
\label{sqrtHol1}
\qqq
\vskip 0.4cm

\noindent Recalling that $\,Hol_\CG(\phi)\,$ was used to define
the WZ amplitudes, \,see (\ref{WZHol}), \,we obtain this way
a local definition of $\,\sqrt{\ee^{\ii S_\WZ(\phi)}}$.

\subsection{$3d$ index with values $\,\pm1$}
\label{subsec:3d_ind}

\noindent Let, as above, $\,\CG\,$ be a gerbe on $\,M\,$ with curvature 
$\,H=\Theta^*H\,$ and let $\,(\eta,\mu)\,$ be a $\,\Theta$-equivariant 
structure on $\,\CG$. Let $\,R\,$ be an oriented $3d$-manifold with an 
orientation-reversing involution $\,\rho\,$ with discrete fixed points and let 
$\,\Phi:R\rightarrow M\,$ satisfy the equivariance condition
\qq
\Phi\circ\rho\,=\,\Theta\circ\Phi\,.
\label{equivPhi}
\qqq   
Let $\,F_R\subset R\,$ be the closure of a fundamental domain for $\,\rho\,$
with smooth boundary $\,\partial F_R\,$ preserved by $\,\rho$.
Define
\qq
\CK(\Phi)\,=\,\frac{\ee^{\frac{\ii}{2}\int_{F_R}\Phi^*H}}{\sqrt{\ee^{
\ii S_\WZ(\Phi|_{\partial F_R})}}}\,.
\qqq
\vskip 0.3cm

\noindent{\bf Proposition 3.} \ Under the same assumption as in Proposition 2,
$\,\CK(\Phi)\,$ is independent of the choice of the fundamental domain $\,F_R\,$
and takes values $\,\pm1$.
\vskip 0.4cm

\noindent That $\CK(\Phi)^2=1$ follows from the definition of the
WZ amplitudes. The proof of independence of the choice of $\,F_R\,$ is done by 
local changes of $\,F_R\,$ and the use of the local construction of  
$\,\sqrt{\ee^{\ii S_\WZ(\Phi|_{\partial F_R})}}\,$ described in 
Sec.\,\ref{subsec:sqrt_hol} above. I do not know how to establish such independence 
without that construction (e.g. employing the definition of the square
root of the WZ amplitude from Sec.\,\ref{sqrtWZ}). It is at this point that 
the bundle gerbe theory shows its utility. Below, we shall consider index 
$\,\CK(\Phi)\,$  for $\,R=\mathbb T^3\,$ with 
$\,\rho\,$ induced by the map $\,k\mapsto -k\,$ in $\,\mathbb R^3$. 
In particular, for the map $\,\mathbb T^3\ni k\mapsto\Phi(k)=I-2P(k)$, 
\,where $\,P(k)=\theta P(-k)\theta^{-1}\,$ are projectors on the valence
band states of a $3d$ time-reversal invariant insulator, \,we shall obtain
the relation:
\qq
\CK(\Phi)=(-1)^{\KM^s}, 
\qqq
where $\,\KM^s\in\mathbb Z_2\,$ is the strong Kane-Mele invariant of such 
insulators \cite{FKM}.

\subsection{Basic gerbe over unitary group} 
\label{subsec:basic_U(N)}

\noindent We would like to apply the above constructions to the case
when $\,M=U(N)\,$ for even $\,N\,$ and $\,\Theta\,$ is induced by the 
adjoint action of the time-reversal anti-unitary operator $\,\theta\,$ 
with $\,\theta^2=-I$.  
\,As already observed, \,there is a unique differential character on the group 
$\,U(N)\,$ corresponding to the closed bi-invariant form $\,H\,$ of 
(\ref{H}). Let us describe here a gerbe $\,\CG=(Y,B,\CL,t)\,$ whose holonomy
gives this character. Such a gerbe, which is unique up to 1-isomorphisms,
is alled basic. The construction presented below is a modified 
version of the one of \cite{MS}. It describes the ambiguities in 
defining a logarithm of unitary matices. One sets
\qq
Y\,=\,\big\{\,(\epsilon,V)\in (-2\pi,0)\times U(N)\,\,\big|\,\,\ee^{-\ii\epsilon}
\not\in spec(V)\,\big\}. 
\qqq
$Y\,$ is then an open subset of $\,(-2\pi,0)\times U(N)\,$ and the projection
$\,\pi\,$ on the second component is clearly a surjective submersion. Then
\qq
Y^{[n]}=\big\{(\epsilon_1,\dots,\epsilon_n,V)\in(-2\pi,0)^n\times U(N)\,\big|\,
\ee^{-\ii\epsilon_i}\notin{spec}(V),\ i=1,\dots,n\big\}.
\qqq
Consider a smooth map from $\,Y\,$ to the Lie algebra $\,u(N)\,$ 
of Hermitian $N\times N$ matrices,
\qq
Y\,\ni\,(\epsilon,V)\,\longrightarrow\,\ii\ln_{-\epsilon}(V)\equiv 
H^{\rm eff}_{\,\epsilon}(V)\,\in\,u(N)\,,
\label{Heff0}
\qqq
where $\,\ln_\phi(r\ee^{\ii\varphi})=\ln r+\ii\varphi\,$ for
$\,\phi-2\pi<\varphi<\phi\,$ is a particular branch of the logarithmic function. 
More explicitly, if
\qq
V=\sum\limits_n\lambda_n\,|\psi_n\rangle\langle\psi_n|
\label{spectV}
\qqq
is the spectral decomposition of $\,V\,$ then
\qq
H^{\rm eff}_{\,\epsilon}(V)=\ii\sum\limits_n\ln_{-\epsilon}(\lambda_n)\,|\psi_n
\rangle\langle\psi_n|\,.
\label{Heff}
\qqq
$H^{\rm eff}_{\,\epsilon}(V)\,$ is a Hermitian matrix
with the spectrum inside the interval $\,(\epsilon,\epsilon+2\pi)$, \,as is
easy to see from the definition of $\,\ln_{-\epsilon}$. \,Moreover, 
\qq
V=\ee^{-\ii H^{\rm eff}_{\,\epsilon}(V)},
\label{baseff}
\qqq
so that $\,V\,$ may be considered as the time-one evolution operator
corresponding to Hamiltonian $\,H^{\rm eff}_{\,\epsilon}$, \,which should
explain the origin of the notation. Note that 
$\,H^{\rm eff}_{\,\epsilon}(V)\,$ is locally constant in $\,\epsilon\,$ 
for fixed $\,V$. \,Let $\,h\,$ be the homotopy
\qq
[0,1]\times Y\,\ni\,(t,\epsilon,V)\,\mathop{\longmapsto}^h\,
(\epsilon,\ee^{-\ii tH^{\rm eff}_{\,\epsilon}(V)})\,\in\,Y\,.
\label{homot2}
\qqq
We shall define the curving 2-form $B$ on $Y$ such that $dB=\pi^*H$
using this homotopy as in the proof of the Poincare Lemma:
\qq
B\,=\,\int_0^1\big(\iota_{\partial_t}h^*\pi^*H\big)\,dt\,.
\label{curving}
\qqq
\vskip 0.2cm

Let us consider the closed 2-form $\,F\,$ on $\,Y^{[2]}$, 
\qq
F\,=\,p_2^*B-p_1^*B\,,\qquad{\rm where}\qquad p_i(\epsilon_1,\epsilon_2,V)
=(\epsilon_i,V)\,,\qquad i=1,2
\label{F1}
\qqq
are two projections from $\,Y^{[2]}\,$ to $\,Y$. 
\,A direct calculation based on the relation 
\qq
H^{\rm eff}_{\epsilon_2}(V)=H^{\rm eff}_{\epsilon_1}(V)+2\pi P_{\epsilon_1,\epsilon_2}(V)\,,
\label{tildeP}
\qqq
holding for $\,\epsilon_1\leq\epsilon_2$, \,where $P_{\epsilon_1,\epsilon_2}(V)$ 
is the spectral projector of $\,V\,$ on the part of the spectrum 
in the sub-interval of the unit circle joining $\,\ee^{-\ii\epsilon_1}\,$ 
to $\,\ee^{-\ii\epsilon_2}\,$ clockwise if $\,\epsilon_1<\epsilon_2\,$ and
$\,P_{\epsilon,\epsilon}=0$, \,gives:
\qq
F(\epsilon_1,\epsilon_2,V)\,=\,\ii\,\tr\,P_{\epsilon_1,\epsilon_2}(V)
(dP_{\epsilon_1,\epsilon_2}(V))^{\wedge 2}\,
-\,\frac{_1}{^4}\,d\,\Big(\tr\,H^{\rm eff}_{\epsilon_1}(V)\,
dP_{\epsilon_1,\epsilon_2}(V)\Big)
\label{F7}
\qqq 
for $\,\epsilon_1\leq\epsilon_2$.
\,We need to construct a line bundle $\,\CL\,$ over $\,Y^{[2]}\,$ with 
curvature $\,F$. \,Let us decompose: 
\qq
Y^{[2]}\,=\,Y^{[2]}_+\sqcup Y^{[2]}_-\sqcup Y^{[2]}_0\,,
\qqq
\vskip -0.4cm
\noindent where
\vskip -0.4cm 
\qq
&&Y^{[2]}_+\,=\,\big\{(\epsilon_1,\epsilon_2,V)\in Y^{[2]}\,\big|\,
\epsilon_1<\epsilon_2,\ P_{\epsilon_1,\epsilon_2}\not=0\big\},\\
&&Y^{[2]}_-\,=\,\big\{(\epsilon_1,\epsilon_2,V)\in Y^{[2]}\,\big|\,
\epsilon_2<\epsilon_1,\ P_{\epsilon_2,\epsilon_1}\not=0\big\},\\
&&Y^{[2]}_0\,=\,\big\{(\epsilon_1,\epsilon_2,V)\in Y^{[2]}\,\big|\,
\epsilon_1\leq\epsilon_2,\ P_{\epsilon_1,\epsilon_2}=0
\ \,{\rm or}\ \,\epsilon_1\geq\epsilon_2,
\ P_{\epsilon_2,\epsilon_1}=0\big\}
\qqq
are disjoint open subsets of $\,Y^{[2]}$. \,Let $\,\sigma\,$ denote the
map $\,(\epsilon_1,\epsilon_2,V)\mapsto(\epsilon_2,\epsilon_1,V)\,$ 
intertwining $\,Y^{[2]}_\pm$. 
There is a tautological vector bundle $\,\CE\,$ (with dimension only locally 
constant, in general) over $\,Y^{[2]}_+\,$ 
whose fiber over $\,(\epsilon_1,\epsilon_2,V)\,$ is $\,P_{\epsilon_1,\epsilon_2}
(V)\,\mathbb C^N$. \,We set
\qq
\CL\big|_{Y^{[2]}_+}\,=\,\det(\CE)\,\equiv\,\wedge^{\rm max}\CE\,,\qquad
\CL\big|_{Y^{[2]}_-}\,=\,\big(\sigma^*\CL|_{Y^{[2]}_+}\big)^{-1}\,,\qquad
\CL\big|_{Y^{[2]}_0}\,=\,Y^{[2]}_0\times\mathbb C\,.
\qqq
Vector bundle $\,\CE\,$ has a Hermitian structure inherited from the 
scalar product in $\,\mathbb C^N\,$ and a unitary connection such that 
for its local section $\,S$,
\qq
\nabla S(\epsilon_1,\epsilon_2,V)\,=\,P_{\epsilon_1,\epsilon_2}(V)\,
dS(\epsilon_1,\epsilon_2,V)\,,
\qqq
Line bundle $\,\CL|_{Y^{[2]}_+}\,$ inherits both these structures 
from $\,\CE$. \,In particular, the induced connection on $\,\CL$, \,that 
we shall denote $\,\nabla^B$, \,is often
called the Berry connection and has the (real) curvature 2-form
\qq
F^B(\epsilon_1,\epsilon_2,V)\,=\,\ii\,\tr\,P_{\epsilon_1,\epsilon_2}(V)
(dP_{\epsilon_1,\epsilon_2}(V))^{\wedge 2}.
\qqq
We shall correct connection $\,\nabla^B\,$ on $\,\CL\big|_{Y^{[2]}_+}\,$ 
adding to it a 1-form $\,-\ii A\,$ on $\,Y^{[2]}_+$, \,where
\qq
A(\epsilon_1,\epsilon_2,V)\,=\,-\frac{_1}{^4}\,\tr\,H^{\rm eff}_{\epsilon_1}(V)\,
dP_{\epsilon_1,\epsilon_2}(V)\,.
\label{Aform}
\qqq 
The curvature of the corrected connection $\,\nabla^B-iA\,$ is then
equal to the 2-form $\,F\,$ of \,(\ref{F7}).
Line bundle $\,\CL|_{Y^{[2]}_-}=\big(\sigma^*\CL|_{Y^{[2]}_+}\big)^{-1}\,$ 
inherits, in turn, the Hermitian structure and the Berry connection 
from $\,\CL|_{Y^{[2]}_+}\,$ and the latter will be corrected by adding 
to it the form $\,\ii\,\sigma^*A$. \,Finally, the trivial line bundle 
$\,\CL_{Y^{[2]}_0}\,$ will be considered with the Hermitian structure 
inherited from $\,\mathbb C\,$ and with the trivial flat connection.
\vskip 0.2cm

We have to define isomorphisms
\qq
t:p_{12}^*\CL\otimes p_{23}^*\CL\,\longrightarrow\,p_{13}^*\CL
\label{tisom}
\qqq
of line bundles over $\,Y^{[3]}$, \,where $\,p_{ab}(\epsilon_1,\epsilon_2,
\epsilon_3,V)=(\epsilon_a,\epsilon_b,V)$. There will be several cases 
of which nontrivial are only the ones when $\,-2\pi<\epsilon_a<\epsilon_b
<\epsilon_c<0\,$ for some permutation of 
$\,(\epsilon_1,\epsilon_2,\epsilon_3)\,$ 
and $\,P_{\epsilon_a,\epsilon_b}(V)\not=0\not=P_{\epsilon_b,\epsilon_c}(V)$. Let 
$\,(u_1,\dots,u_k)\,$ be an orthonormal basis of the range of 
$\,P_{\epsilon_a,\epsilon_b}(V)\,$ 
and $\,(u_{k+1},\dots,u_{k+l})\,$ an orthonormal basis of the range of 
$P_{\epsilon_b,\epsilon_c}(V)$. 
Then $\,(u_1,\dots,u_{k+l})\,$ is an orthonormal basis 
of the range of $P_{\epsilon_a,\epsilon_c}(V)$. We shall denote by 
$\,(u_1\wedge\cdots\wedge u_k)^{-1}\,$ the element of the dual line
$\,\big(\wedge^{\rm max}P_{\epsilon_a,\epsilon_b}(V)\,\mathbb C^N\big)^{-1}\,$ that 
pairs to $\,1\,$ with $\,u_1\wedge\cdots\wedge u_k$, \,etc., \,and \,set
\qq
&&\hbox to 9.4cm{$t\big((u_1\wedge\cdots
\wedge u_k)\otimes(u_{k+1}\wedge\cdots\wedge u_{k+l})\big)\,=\,
u_1\wedge\cdots\wedge u_{k+l}$\hfill}\qquad{\rm for}\ \quad\epsilon_1<
\epsilon_2<\epsilon_3\,,
\label{case123}\\
&&\hbox to 9.4cm{$t\big((u_{k+1}\wedge\cdots
\wedge u_{k+l})^{-1}\otimes(u_1\wedge\cdots\wedge u_k)^{-1}\big)
\,=\,(u_1\wedge\cdots\wedge u_{k+l})^{-1}$\hfill}\qquad{\rm for}\ \quad 
\epsilon_3<\epsilon_2<\epsilon_1\,,\label{case321}\\
&&\hbox to 9.4cm{$t\big((u_1\wedge\cdots
\wedge u_{k+l})\otimes(u_{k+1}\wedge\cdots\wedge u_{k+l})^{-1}\big)\,=\,
u_1\wedge\cdots\wedge u_{k}$\hfill}\qquad{\rm for}\ \quad \epsilon_1
<\epsilon_3<\epsilon_2\,,
\label{case132}\\
&&\hbox to 9.4cm{$t\big((u_{k+1}\wedge\cdots
\wedge u_{k+l})\otimes(u_1\wedge\cdots\wedge u_{k+l})^{-1}\big)\,=\,
(u_1\wedge\cdots\wedge u_{k})^{-1}$\hfill}\qquad{\rm for}\ \quad \epsilon_3
<\epsilon_1<\epsilon_2\,,
\label{case312}\\
&&\hbox to 9.4cm{$t\big((u_1\wedge\cdots
\wedge u_k)^{-1}\otimes(u_1\wedge\cdots\wedge u_{k+l})\big)\,=\,
u_{k+1}\wedge\cdots\wedge u_{k+l}$\hfill}\qquad{\rm for}\ \quad\epsilon_2
<\epsilon_1<\epsilon_3\,,
\label{case213}\\
&&\hbox to 9.4cm{$t\big((u_1\wedge\cdots
\wedge u_{k+l})^{-1}\otimes(u_1\wedge\cdots\wedge u_k)\big)\,=\,
(u_{k+1}\wedge\cdots\wedge u_{k+l})^{-1}$\hfill}\qquad{\rm for}\ \quad 
\epsilon_2<\epsilon_3<\epsilon_1\,.\label{case231}
\qqq
It is easy to see that the isomorphism $\,t\,$ defined this way intertwines 
the Hermitian structures and the Berry connections. The following lemma 
shows that it intertwines also the connections modified by the addition 
of 1-forms (\ref{Aform}):  
\vskip 0.4cm

\noindent{\bf Lemma.}\ \ For $\,(\epsilon_1,\epsilon_2,\epsilon_3,V)
\in Y^{[3]}\,$ and $\,-2\pi<\epsilon_a<\epsilon_b<\epsilon_c<0$,
\qq
A(\epsilon_a,\epsilon_b,V)+A(\epsilon_b,\epsilon_c,V)
=A(\epsilon_a,\epsilon_c,V)\,.
\label{Aadd}
\qqq
\vskip 0.3cm

It is easy, although somewhat tedious, to check that the
isomorphism $\,t\,$ is associative. Such check completes the construction 
of a basic gerbe $\CG=(Y,B,\CL,t)\,$ over $\,U(N)$.
\vskip 0.3cm

\subsection{Towards time-reversal equivariant basic
gerbe over $\,U(N)$}
\label{subsec:Theta_equiv}

\noindent Recall that $\,\Theta\,$ acts on $\,U(N)\,$ for $\,N\,$ 
even by conjugation with the anti-unitary map $\,\theta\,$ of $\,\mathbb C^N$
such that $\,\theta^2=-I$. 
We would like to construct a $\,\Theta$-equivariant structure on the
basic gerbe $\,\CG\,$ over $\,U(N)$. In the first step, \,we need to obtain 
a 1-isomorphism $\,\eta:\CG\rightarrow\Theta^*\CG$. \,Define the involution
$\,\Theta_Y:Y\rightarrow Y\,$ covering $\,\Theta\,$ by
\qq
\Theta_Y(\epsilon,V)\,=\,(-\epsilon-2\pi,\Theta(V))\,.
\label{ThetaY}
\qqq
The pullback gerbe $\Theta^*\CG\,$ is represented by the quadruple 
$\,(Y,\Theta_Y^*B,(\Theta_Y^{[2]})^*\CL,(\Theta_Y^{[3]})^*t)$.
If the spectral decomposition of $\,V\,$ is given by (\ref{spectV}) then
\qq
\Theta(V)=\sum\limits_n\overline{\lambda_n}\,
\theta|\psi_n\rangle\langle\psi_n|\theta^{-1}
\qqq
is the spectral decomposition of $\,\Theta(V)$. \,In particular, the spectrum
of fixed points of $\,\Theta\,$ is symmetric under 
$\,\lambda\mapsto\bar\lambda$. \,Relation
$\,\ln_{-\epsilon}(\lambda)=-\ln_{\epsilon+2\pi}(\overline{\lambda})\,$ implies that:
\qq
\theta H^{\rm eff}_{\,\epsilon}(V)\,\theta^{-1}\,=\,\ii\sum_n\ln_{-\epsilon}(\lambda_n)\,
\theta|\psi_n\rangle\langle\psi_n|\theta^{-1}\,=\,-\ii\sum\limits_n
\ln_{\epsilon+2\pi}(\overline{\lambda_n})\,\theta|\psi_n\rangle
\langle\psi_n|\theta^{-1}
\,=\,-H^{\rm eff}_{-\epsilon-2\pi}(\Theta(V))\,.\quad
\label{thetaeff}
\qqq
Using this identity, it is straightforward to check that
$\,\Theta_Y^*B=B$.
\vskip 0.1cm

Now, it is easy to see that line bundles $\,\CL\,$ and $\,(\Theta_Y^{[2]})^*\CL\,$
over $Y^{[2]}$ are isomorphic. Since
\qq
\Theta_Y^{[2]}(\epsilon_1,\epsilon_2,V)=(-\epsilon_1-2\pi,-\epsilon_2-2\pi,
\Theta(V))\,,
\qqq
it follows that $\,\Theta_Y^{[2]}\,$ intertwines the subsets $\,Y^{[2]}_\pm\,$ 
and leaves $\,Y^{[2]}_0\,$ invariant. If $\,(\epsilon_1,\epsilon_2,V)
\in Y^{[2]}_+\,$ and $\,(u_1,\dots,u_k)\,$ is an orthonormal basis of the range 
of $\,P_{\epsilon_1,\epsilon_2}(V)\,$ then $\,(\theta u_1,\dots,\theta u_k)\,$ 
is an orthonormal basis of the range of $\,P_{-\epsilon_2-2\pi,-\epsilon_1-2\pi}
(\Theta(V))$. Similarly, if $\,(\epsilon_1,\epsilon_2,V)\in Y^{[2]}_-\,$ and 
$\,(u_1,\dots,u_k)\,$ is an orthonormal bases 
of the range of $\,P_{\epsilon_2,\epsilon_1}(V)\,$ then $\,(\theta u_1,\dots,
\theta u_k)\,$ is an orthonormal basis of 
$\,P_{-\epsilon_1-2\pi,-\epsilon_2-2\pi}(\Theta(V))$.
The line bundle isomorphism $\,\nu:\CL\rightarrow(\Theta_Y^{[2]})^*\CL\,$ 
(linear on fibers!) defined by
\qq
&&\nu(u_1\wedge\cdots\wedge u_k)=(\theta u_k\wedge\cdots\wedge\theta u_1)^{-1}
\qquad{\rm for}\qquad(\epsilon_1,\epsilon_2,V)\in Y^{[2]}_+\,,
\label{wedgetheta1}\\     
&&\nu(u_1\wedge\cdots\wedge u_k)^{-1}=(\theta u_k\wedge\cdots\wedge\theta u_1)
\qquad{\rm for}\qquad(\epsilon_1,\epsilon_2,V)\in Y^{[2]}_-\,,
\label{wedgetheta2}
\qqq
and by identifying the trivial bundles over $\,Y^{[2]}_0$, \,intertwines 
the Hermitian structures and the Berry connections. It also intertwines 
the corrected connections as follows from the the relation 
\qq
(\Theta_Y^{[2]})^*(-\sigma^*A)=A
\qqq
that holds on $\,Y^{[2]}_+$. 
It is straightforward to check that $\,\nu\,$ intertwines the groupoid
multiplications $\,t\,$ and $\,(\Theta^{[2]}_Y)^*t$ \,(the change of the order
of vectors in (\ref{wedgetheta1}) and (\ref{wedgetheta2}) is essential here). 
We infer that isomorphism $\,\nu\,$ together with a trivial bundle $\,\CN\,$ 
over $\,Y\,$ defines a 1-isomorphism $\,\eta:\CG\rightarrow\Theta^*\CG$.
\vskip 0.2cm

Let us observe that 
\qq
(\Theta_Y^{[2]})^*\nu\circ\nu\,=\,(-1)^\kappa\Id_\CL\,,
\label{Tnn}
\qqq
\noindent where $\,\kappa(\epsilon_1,\epsilon_2,V)\,$ is equal to the dimension
of the range of $\,P_{\epsilon_1,\epsilon_2}(V)\,$ on $\,Y^{[2]}_+\,$ and of
$\,P_{\epsilon_2,\epsilon_1}(V)\,$ on $\,Y^{[2]}_-\,$
and vanishes on $\,Y^{[2]}_0$. \,The sign in (\ref{Tnn}) is induced by 
the action of $\,\theta^2=-I$. 
\,As discussed in Sec.\,\ref{subsec:equivariant_gerbes}, 1-isomorphism 
$\,\Theta^*\eta\circ\eta:\CG\rightarrow\CG\,$ corresponds
to a flat line bundle $\,Q\,$ over $\,U(N)$. \,We may take  
\qq
Q\,=\,(Y\times\mathbb C)\big/\sim
\label{lbQ}
\qqq
where $\,\sim\,$ is the equivalence relation
\qq
(\epsilon_1,V,z_1)\,\sim\,(\epsilon_2,V,z_2)\qquad{\rm if}\qquad 
z_2=(-1)^{k(\epsilon_1,\epsilon_2,V)} z_1\,.
\label{equivr}
\qqq
Elements of $\,Q\,$ will be denoted $[\epsilon,V,z]_{_\sim}$. 
\,Projection $\,\pi_Q:Q\rightarrow U(N)\,$ forgets $\,\epsilon\,$ and $\,z$.
\,Line bundle $\,Q\,$ comes with the involution $\,\Theta_Q$, 
\qq
\Theta_Q([\epsilon,V,z]_{_\sim})\,=\,([-\epsilon-2\pi,\Theta(V),z]_{_\sim}\,,
\label{ThetaQ}
\qqq
that is linear on the fibers and covers involution $\,\Theta$. 
\vskip 0.1cm
 
Now, 1-isomorphism $\,\Theta^*\eta\circ\eta\,$ is 2-isomorphic
to a trivial one if and only if the flat line bundle $\,Q\,$ is trivializable.
This holds if and only if the holonomy of $\,Q\,$ along the loop 
\qq
[0,2\pi]\ni\varphi\,\longmapsto\,V(\varphi)
={\rm diag}(\ee^{\ii\varphi},1,\dots,1)
\label{loopV}
\qqq
in $\,U(N)\,$ is equal to $\,1$. Indeed, the above loop is a generator of 
$\,\pi_1(U(N))\cong\mathbb Z\,$ (the latter isomorphism is given by the winding 
number of the determinant). Consider
\qq
W(\varphi)\,=\,\begin{cases}\,[-\frac{_{4\pi}}{^3},V(\varphi),1]_{_\sim}\qquad
{\,\rm if}\qquad \varphi\in[0,\pi]\,,\cr
\,[-\frac{_{2\pi}}{^3},V(\varphi),-1]_{_\sim}\quad\hspace{0.14cm}
{\rm if}\qquad \varphi\in[\pi,2\pi]\,.
\end{cases}
\qqq
$[0,2\pi]\ni\varphi\mapsto W(\varphi)\,$ is a horizontal lift to $\,Q\,$
of the loop $\,V(\phi)$. \,Since $\,W(2\pi)=-W(0)$, \,it follows that 
the holonomy of $\,Q\,$ along the loop (\ref{loopV}) is equal to $\,-1\,$
so that $\,Q\,$ is not trivializable.
Consequently, \,1-isomorphism $\,\Theta^*\eta\circ\eta:
\CG\rightarrow\CG\,$ is not 2-isomorphic to the identity 1-isomorphism. 
A choice of a different 1-isomorphism $\,\eta:\CG\rightarrow\CG\,$ would
change the flat line bundle $\,Q\,$ by a tensor factor 
$\,\Theta^*\tilde Q\otimes\tilde Q\,$ for some flat bundle $\,\tilde Q$ 
over $\,U(N)\,$ and would not change the holonomy along loop 
(\ref{loopV}). We conclude that there is no $\,\Theta$-equivariant 
structure on the basic gerbe $\,\CG\,$ on $\,U(N)$! \,Note however that
the holonomy of flat line bundle $\,Q\,$ squares to $\,1$. 
\vskip 0.4cm

\noindent{\bf Remark.} \ In the case when $\theta^2=I$, \,the same construction
gives 1-isomorphism $\,\Theta^*\eta\circ\eta\,$ coinciding with $\,\Id_{\CG}\,$
so that the $\,\Theta$-equivariant structure on $\,U(N)\,$ does 
exist in that case.

\subsection{Lift to the double cover of $\,U(N)$}
\label{subsec:lift}

\noindent A possible way out from the difficulty encountered in the previous
section is to pass to the double cover $\,\widehat U(N)\,$ 
of \,group $\,U(N)$, 
\qq
\widehat U(N)\,=\,\big\{(V,\omega)\in U(N)\times U(1)\,\big|\,
\omega^2=\det(V)\big\}
\label{2ndreal}
\qqq
with covering map $\,\widehat\pi:\widehat U(N)\rightarrow U(N)\,$ forgetting
$\,\omega$. Let us denote by $\,\widehat\Theta\,$ the involution
\qq
(V,\omega)\,\mathop{\longmapsto}\limits^{\widehat\Theta}\,(\Theta(V),\omega^{-1})
\label{widehatTheta}
\qqq
of $\,\widehat U(N)$, \,by $\,\widehat Q\,$
the pullback $\,\widehat\pi^*Q\,$ of line bundle $\,Q\,$ of
(\ref{lbQ}) to $\,\widehat U(N)\,$
and by $\,\widehat\Theta_{\widehat Q}\,$ the pullback $\,\widehat\pi^*\Theta_Q\,$
of the involution $\,\Theta_Q\,$ of (\ref{ThetaQ}) that is explicitly given
by the relation
\qq
\widehat\Theta_{\widehat Q}\big((V,\omega),[\epsilon,V,z]_{_\sim}\big)=
\big((\Theta(V),\omega^{-1}),[-\epsilon-2\pi,\Theta(V),z]_{_\sim}\big)\,.
\qqq
$\widehat\Theta_{\widehat Q}\,$ is an involution of $\,\widehat Q\,$
linear on fibers that covers $\,\widehat\Theta$.
\vskip 0.1cm

Unlike $\,Q$, \,the flat line bundle $\,\widehat Q\,$ possesses global
section $\,\widehat S\,$ given by the formula
\qq
\widehat S(V,\omega)\,
=\,((V,\omega),[\epsilon,V,\,\omega\,\det(\ee^{\frac{\ii}{2}H^{\rm eff}_{\,\epsilon}(V)})]_{_\sim})\,.
\qqq
An easy check based on relation (\ref{tildeP}) shows that the equivalence class
on the right hand side is well defined. 
Since $\,\big(\omega\,\det(\ee^{\frac{\ii}{2}H^{\rm eff}_{\,\epsilon}(V)})
\big)^2=1$, \,it follows that section $\,\widehat S\,$ is flat and normalized.
Besides
\qq
\widehat\Theta_{\widehat Q}
\circ\widehat S(V,\omega)&=&
((\Theta(V),\omega^{-1}),[-\epsilon-2\pi,\Theta(V),\,\omega\,
\det(\ee^{\frac{\ii}{2}H^{\rm eff}_{\,\epsilon}(V)})]_{_\sim})\cr
&=&((\Theta(V),\omega^{-1}),[-\epsilon-2\pi,\Theta(V),\,\omega^{-1}\,
\det(\ee^{\frac{\ii}{2}H^{\rm eff}_{\,\epsilon}(V)})^{-1}]_{_\sim})\cr
&=&((\Theta(V),\omega^{-1}),[-\epsilon-2\pi,\Theta(V),\,\omega^{-1}\,
\det(\ee^{\frac{\ii}{2}H^{\rm eff}_{-\epsilon-2\pi}(\Theta(V))})]_{_\sim})\,=\,
S(\Theta(V),\omega^{-1})\,,
\label{pent}
\qqq 
where the last but one equality is a consequence of (\ref{thetaeff}).
\vskip 0.1cm
 
It follows that, although there is no $\,\Theta$-equivariant structure 
on the basic gerbe $\,\CG\,$ over $\,U(N)$, there is a 
$\,\widehat\Theta$-equivariant structure on the pullback gerbe 
$\,\widehat\CG=\widehat\pi^*\CG\,$ over
$\,\widehat U(N)$. \,Indeed, $\,\widehat\eta=\widehat\pi^*\eta\,$
provides a 1-isomorphism $\,\widehat\eta:\widehat\CG\rightarrow
\widehat\pi^*\Theta^*\CG=\widehat\Theta^*\widehat\CG\,$
and the trivializing section $\,\widehat S\,$ 
of the flat line bundle $\,\widehat Q\,$ defines a 2-isomorphisms 
$\,\widehat\mu:\widehat\Theta^*\widehat\eta\circ\widehat\eta\Rightarrow
\Id_{\widehat\CG}$. \,Relation (\ref{pent}) assures that condition 
(\ref{mu2}) holds for $\,\widehat\mu$.
\vskip 0.2cm

The set $\,U(N)'$ of $\,\Theta$-invariant points of $\,U(N)\,$ is a 
closed subgroup conjugate to group $\,Sp(N)$, which is connected
and simply connected. In particular, $\,\det(V)=1\,$ for $\,V\in U(N)'$.
\,This implies that the subset $\,\widehat U(N)'\,$ of 
$\,\widehat\Theta$-invariant points in $\,\widehat U(N)\,$ coincides 
with the subgroup $\,U(N)'\times\{\pm1\}\,$ that is simply connected but 
has two connected components that we shall accordingly denote 
$\,\widehat U(N)'_\pm\cong U(N)'$. 
\,It follows that the restriction of section 
$\,\widehat S'=\widehat S|_{\widehat U(N)'}\,$ 
to $\,\widehat U(N)'_\pm\,$ may be identified with the sections
$\,S'_{\pm}\,$ of bundle $\,Q'=Q|_{U(N)'}\,$ defined by
\qq
U(N)'\ni V\,\mathop{\longmapsto}^{S'_\pm}\ 
[\epsilon,V,\pm\det(\ee^{\frac{\ii}{2}H^{\rm eff}_{\,\epsilon}})]_{_\sim}\,.
\label{Spm}
\qqq
\vskip 0.2cm
   
\begin{figure}[!h]
\begin{center}
\leavevmode
 \includegraphics[width=5.7cm,height=3.8cm]{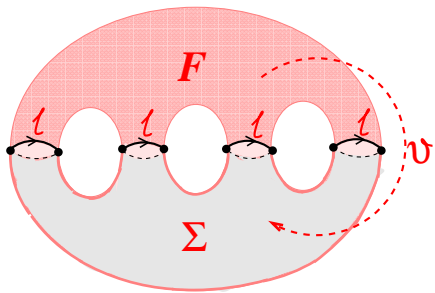}
\vskip 0.1cm
\caption{Closed surface $\,\Sigma\,$ of genus $\,g=3\,$ 
with fundamental domain $\,F\,$ for orientation preserving involution 
$\,\vartheta\,$ with $\,8=2g+2\,$ fixed points marked as black dots}
\label{fig:sigmaF}
\end{center}
\vskip -0.3cm
\end{figure}

Suppose now that $\,\phi:\Sigma\mapsto U(N)\,$ is an equivariant map,
\,i.e. that relation (\ref{thetaphi}) holds for it.
For $\,\Sigma\,$ which is a 2-torus or a hyperelliptic curve
with involutions $\,\vartheta\,$ described before, \,one may choose a 
base of 1-homology of $\,\Sigma\,$ composed of loops that are invariant 
under $\,\vartheta$, \,up to the orientation change, and each containing 
two fixed points of $\,\vartheta$, \,see Fig.\,\ref{fig:sigmaF}. 

\noindent Determinant of field $\,\phi\,$ satisfies the relation 
\qq
\det(\phi(\vartheta x))=\det(\Theta\circ\phi(x))=\det(\phi(x))^{-1}
\qqq
and $\,\det(\phi(x))=1\,$ at fixed points of $\,\vartheta$. \,It follows
that $\,\det\circ\phi\,$ has even winding numbers around the homology
1-cycles of $\,\Sigma$. As the result, $\,\phi\,$ may be lifted to
a map $\,\widehat\phi:\Sigma\rightarrow\widehat U(N)$. \,The lift
is unique up to a multiplication by $\,(I,-1)\in\widehat U(N)$.
\,Besides, it is still equivariant:
\qq
\widehat\phi\circ\vartheta=\widehat\Theta\circ\widehat\phi
\qqq
because this property holds at fixed points of $\,\vartheta$.
We may then define
\qq
\sqrt{\ee^{\ii S_{\WZ}(\phi)}}\,=\,
\sqrt{\ee^{\ii S_{\WZ}(\widehat\phi)}}\,=\,\sqrt{Hol_{\widehat\CG}(\widehat\phi)}\,,
\label{sqrthatS}
\qqq
where the right hand side is determined using the $\widehat\Theta$-equivariant
structure on $\,\widehat U(N)$, \,see Sec.\,\ref{subsec:sqrt_hol}. Some care 
is needed, however.
\vskip 0.1cm

First, the subset $\,\widehat U(N)'\,$ of fixed points of 
$\,\widehat\Theta$, \,although simply connected, is disconnected. 
As the result, we may multiply the trivializing 
section $\,\sqrt{\widehat S'}\,$ of line bundle 
$\,\widehat N'\,$ (that is the pullback to $\,\widehat U(N)'$ of line 
bundle $\,N'$ over $\,U(N)'$) \,by $\,\pm1\,$ chosen differently on 
two components of $\,\widehat U(N)'$. 
For vertices $\,v\,$ at the two ends of a connected component of 
$\,\ell\subset\partial F$, \,see Fig.\,\ref{fig:sigmaF}, the lifted 
field $\,\widehat\phi\,$ takes 
values in the same component of $\,\widehat U(N)'\,$ if $\,\det\circ\phi\,$ 
winds an even number of times around zero along that component and 
in different components if it winds an odd number of times. Hence, 
the changes of sign in section $\,\sqrt{\widehat S'}\,$ described above 
result in the multiplication of $\,\sqrt{Hol_{\widehat\CG}(\widehat\phi)}\,$
defined by (\ref{sqrtHol1}) by $\,(-1)^w\,$ where $\,w\,$ is the total 
winding number of $\,\det\circ\phi\,$ along $\,\ell$. \,Note, however, 
that $\,2w\,$ is the total winding number of $\,\det\circ\phi\,$
along the boundary $\,\partial F\,$ of the fundamental domain
$\,F\subset\Sigma\,$ and the latter is necessarily equal to zero.
We infer that even if the fixed point set $\,\widehat U(N)'\;$ is not 
connected, $\,\sqrt{Hol_{\widehat\CG}(\widehat\phi)}\,$
defined by (\ref{sqrtHol1}) does not depend on the choice of section 
$\,\sqrt{\widehat S'}$. 
\vskip 0.1cm

Second, we should show that $\,\sqrt{Hol_{\widehat\CG}(\widehat\phi)}\,$ 
does not depend on the choice of the lift $\,\widehat\phi\,$
of $\,\phi\,$ to $\,\widehat U(N)$. \,Indeed, 
$\,\sqrt{Hol_{\widehat\CG}(\widehat\phi)}\,$  calculated for a different 
lift is equal to the one for the original lift but obtained using
the $\,\widehat\Theta$-equivariant structure of $\,\widehat\CG\,$ 
with 2-isomorphism $\,\widehat\mu\,$ corresponding to section 
$\,-\widehat S\,$ of line bundle $\,\widehat Q\,$ and to section 
$\,\ii\sqrt{\widehat S'}\,$ of line bundle $\,\widehat N'$. 
\,Such modification does not change the phase associated 
to the left hand side of (\ref{sqrtHol1}) as the additional factors
in $\,\sqrt{\widehat S'}(\widehat\phi(v))^{\pm1}\,$ from the ends of each 
connected component of $\,\ell\,$ cancel each other (note that this 
would be also the case if we multiplied $\,\widehat S\,$ by any phase). 
As the result, square root of the WZ amplitudes of equivariant maps 
$\,\phi\,$ is uniquely defined by formula (\ref{sqrthatS}) for the 
special type of surfaces with involution $\,(\Sigma,\vartheta)\,$ 
considered above.  
\vskip 0.4cm

\noindent{\bf Remark.} \,The construction described above would not work
for surfaces with involution like in Fig.\,\ref{fig:sigmaF} but with
an additional handle inside $\,F\,$ and its $\,\vartheta$-image.
\vskip 0.4cm

We shall need a more explicit description of section
$\,\sqrt{\widehat S'}\,$ trivializing the flat line bundle $\,\widehat N'$. 
Let us look first at the line bundle $\,N'\,$ over $\,U(N)'$
whose pullback to $\,Y'\,$ may be identified with the bundle
$\,\CN'$ of (\ref{CN'}). We shall suppose below that 
$\,-\epsilon-2\pi<\epsilon$. \,Then
\qq
\CN'_{(\epsilon,V)}\,=\,\CL_{(-\epsilon-2\pi,\epsilon,V)}\otimes
\CN_{(\epsilon,V)}\,=\,
\,\wedge^{\rm max}P_{-\epsilon-2\pi,\epsilon}(V)\,\mathbb C^N
\qqq
for $\,V\in U(N)'\,$ and the bundle isomorphism 
$\,\nu':\CN'\rightarrow \Theta_Y^*\CN'\,$ is given by
\qq
\hspace{0.1cm}\CN'_{(\epsilon,V)}\ni(u_1\wedge\cdots\wedge u_m)\,\longmapsto\,
(\theta u_m\wedge\cdots\wedge\theta u_1)^{-1}\in \CN'_{(-\epsilon-2\pi,V)}\,, 
\qqq
where $\,(u_1,\dots,u_m)\,$ is an orthonormal basis of the range of
$\,P_{-\epsilon-2\pi,\epsilon}(V)\,$ ($m\,$ is necessarily even). The map 
(\ref{CN'2}) takes the form
\qq
\CN'_{\epsilon,V}\otimes\CN'_{\epsilon,V}\ni(u_1\wedge\cdots\wedge u_m)\otimes
(u_1\wedge\cdots\wedge u_m)&\longmapsto&
(\theta u_m\wedge\cdots\wedge\theta u_1)^{-1}\otimes
(u_1\wedge\cdots\wedge u_m)\cr\cr
&\longmapsto&(-1)^{m/2}\,
\det(\langle u_i|\theta u_j\rangle)^{-1}\,.
\label{CN'22}
\qqq
and induces the line bundle isomorphism $\,N'\otimes N'\rightarrow Q'$.
\,The restriction of the section $\,\sqrt{\widehat S'}\,$
of line bundle $\,\widehat N'\,$ to $\,\widehat U(N)'_\pm\,$
may be identified with the section $\,\sqrt{S'_\pm}\,$ of
line bundle $\,N'\,$ obtained from the assignment
\qq
U(N)'\ni V\,\longmapsto
\,\sigma\,{\rm pf}(\langle u_i|\theta u_j\rangle)
\,u_1\wedge\cdots\wedge u_m\in\CN'_{(\epsilon,V)}\,,
\label{S'pm1/2}
\qqq
where $\,\sigma^2=\pm1\,$ for $\,\sqrt{S'_\pm}\,$ and $\,{\rm pf}(\,\cdot\,)\,$
stands for the Pfaffian of an antisymmetric matrix. It is somewhat tedious
but straightforward to check that, indeed, (\ref{S'pm1/2}) defines a section 
$\,\sqrt{S'_\pm}\,$ of $\,N'$. \,Besides, under the line bundle isomorphism 
$\,N'\otimes N'\rightarrow Q'\,$ induced by (\ref{CN'22})
\qq
\sqrt{S'_\pm}(V)\otimes\sqrt{S'_\pm}(V)\,\longmapsto\,
[\epsilon,V,\pm(-1)^{m/2}]_{_\sim}\,=\,S'_\pm(V)\,,
\qqq
where the last equality follows from (\ref{Spm}) and the relation
$\,\det(\ee^{\frac{\ii}{2}H^{\rm eff}_{\,\epsilon}(V)})=(-1)^{m/2}\,$ holding for 
$\,V\in U(N)'$ that is easy to check. Formula (\ref{S'pm1/2}) provides 
an explicit description of the trivializations $\,\sqrt{S'_\pm}\,$ of 
line bundle $\,N'\,$ and, in the final count, of 
$\,\sqrt{Hol_{\widehat \CG}(\widehat\phi)}\,$ which ends up as given 
by the formula
\qq
Hol_\CG(\phi|_F)\otimes hol_{\CN}(\phi|_\ell)\,=\,
\sqrt{Hol_{\widehat\CG}(\widehat\phi)}\mathop{\otimes}\limits_{v\in\partial\ell}
\sqrt{S'_\pm}(\phi(v))^{\pm1}\,,
\label{sqrtHol2}
\qqq
where the signs in $\,\sqrt{S'_\pm}\,$ have to be chosen consistently with
the winding of $\,\det\circ\phi$, \,as discussed above. This is the only
modification with respect to formula (\ref{sqrtHol1}).
\vskip 0.2cm

It is not difficult to show by studying the evolution of the left hand 
side of (\ref{sqrtHol1}) under smooth changes of $\,\phi\,$ that, 
for $\,\Sigma=\mathbb T^2\,$  with $\,\vartheta k=-k$, \,the square root 
of the WZ amplitude defined by (\ref{sqrthatS}) coincides with the one 
defined in Sec.\,\ref{subsec:WZamplis} via a $3d$ integral.
\vskip 0.2cm

In a similar way as for the square root of the WZ amplitude discussed above, 
one may define the $3d$ index $\,\CK(\Phi)\,$ of an equivariant
map to $\,U(N)\,$ from a $3d$ torus with the orientation reversing 
involution induced by the map $\,k\mapsto -k$. One just sets $\,\CK(\Phi)=
\CK(\widehat\Phi)$, \,where $\,\widehat\Phi\,$ is one of the two lifts of 
$\,\Phi\,$ to $\,\widehat U(N)$. \,The independence on the choice of the 
lift follows from the similar property for the square root of WZ amplitudes
on $2d$ tori, at least when the fundamental domain $\,F_R\,$ chosen for
the calculation of $\,\CK(\widehat\Phi)\,$ has the boundary composed of
two 2-dimensional tori. We shall extensively use such $3d$ index in 
Lecture 2. 
\vskip 0.4cm

\noindent{\bf Remark.} \ Although the time-reversal symmetry with
$\,\theta^2=I\,$ gives rise to a $\,\Theta$-equivariant structure on the
basic gerbe $\,\CG\,$ over $\,U(N)$, \,the fixed point subgroup $\,U(N)'$
is conjugate to group $\,O(N)\,$ here and is neither connected nor simply 
connected. The square root of the WZ amplitudes of equivariant maps 
is not well defined in this case.

\nsection{Lecture 2. \,Applications to topological insulators}
\label{sec:lect2}

\subsection{Crystalline systems and Bloch theory}

\noindent We shall be interested in properties of condensed matter 
systems with crystalline symmetry. The simplest models of such systems
have the space of states 
\qq
\CH\,=\,L^2(\CC,\mathbb V)\,,
\qqq 
where $\,\CC$, a ``crystal'', is an infinite discrete subset of 
$\,d$-dimensional Euclidean space $\,\mathbb E^d\,$ symmetric under 
a group $\,\Gamma\cong\mathbb Z^d\,$ of discrete translations 
(the ``Bravais lattice'') and $\,\mathbb V\,$ is a finite-dimensional Hilbert 
space of internal degrees of freedom (like spin). 
\qq
\Gamma=\big\{\,\sum\limits_{i=1}^d n_ia_i\ \big|\ n_i\in\mathbb Z\,\big\},
\qqq
where $\,a_i\,$ are $\,d\,$ linearly independent vectors in $\,\mathbb R^d$. 
\begin{figure}[h]
\leavevmode
\begin{center}
\includegraphics*[width=5.5cm,height=4.6cm,angle=0]{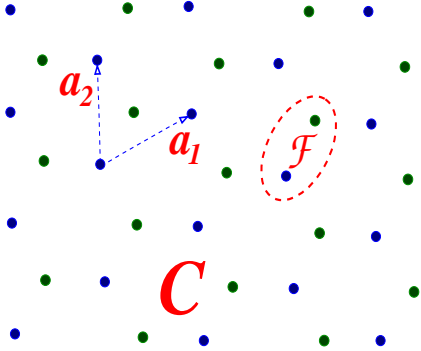}\\
\end{center}
\caption{The hexagonal crystal} 
\label{fig:crystal}
\vskip 0.2cm
\end{figure}
\,The action of $\,\Gamma\,$ on $\,\CC\,$ induces a representation 
of $\,\Gamma\,$ in the space of states by the formula
$\,(U_a\psi)(x)=\psi(x-a)\,$ for $\,x\in\CC\,$ and $\,a\in\Gamma$.
\,The Fourier transform $\,\psi\mapsto\hat\psi\,$ over $\,\Gamma$,
\qq
\hat\psi_k(x)\,=\,\sum\limits_{a\in\Gamma}\ee^{-\ii k\cdot a}\,\psi(x-a)
\qqq
for $\,k\in\mathbb R^d$, \,takes values in Bloch functions satisfying 
the twisted periodic conditions 
$\,\hat\psi_k(x-a)=\ee^{\ii k\cdot a}\,\hat\psi_k(x)$.
\,Note that $\,\psi_k(x)=\psi_{k+b}(x)\,$ for 
$\,b\in 2\pi\Gamma^*\equiv\Gamma^r$, \,the ``reciprocal lattice''. 
The inverse formula states that
\qq
\psi(x)\,=\,\frac{_1}{^{|\BZ|}}\int\limits_{\BZ}\hat\psi_k(x)\,dk\,,
\qqq
where $\,\BZ=\mathbb R^d/\Gamma^r\cong\mathbb T^d\,$ is the 
``Brillouin torus'' and $\,|\BZ|\,$ denotes its Euclidean volume.
The set of Bloch functions for fixed $\,k\,$ forms a finite-dimensional
Hilbert space $\,\hat\CH_k=\hat\CH_{k+b}\,$ with the norm squared
\qq
\Vert\hat\psi_k\Vert^2\,=\,\sum\limits_{x\in\CF}
|\hat\psi_k(x)|^2 
\qqq
where $\,\CF\subset\CC\,$ is an arbitrary unit cell of the crystal having 
exactly 1 representative in each coset of $\,\CC/\Gamma$, 
\,see Fig.\,\ref{fig:crystal}. The Plancherel formula states that 
\qq
\sum\limits_{x\in\CC}|\psi(x)|^2\,=\,\frac{_1}{^{|\BZ|}}\int\Vert\hat\psi_k\Vert^2
\,dk\,.
\qqq
Restriction of Bloch functions to a fixed unit cell $\,\CF\,$ provides 
an identification $\,\CH_k\cong L^2(F,\mathbb V)\cong\mathbb C^N$, where
$\,N=|\CF|\times{\rm dim}(\mathbb V)$. Hence the Fourier transform may 
be viewed as
a unitary isomorphism 
\qq
L^2(\CC,\mathbb V)\,\cong\,L^2(\BZ,\mathbb C^N)
\label{FTr}
\qqq
sending $\,\mathbb V$-valued functions $\,\psi(x)\,$ on $\,\CC\,$ to $\,\mathbb 
C^N$-valued functions $\,\hat\psi(k)\,$ on $\,\BZ$. Compactly supported 
functions on $\,\CC\,$ correspond to analytic functions on $\,\BZ\,$ and 
the action of $\,U(a)\,$ becomes the multiplication by $\,\ee^{\ii k\cdot a}$. 
\vskip 0.2cm

\begin{figure}[h]
\leavevmode
\begin{center}
\includegraphics*[width=7cm,height=4.8cm,angle=0]{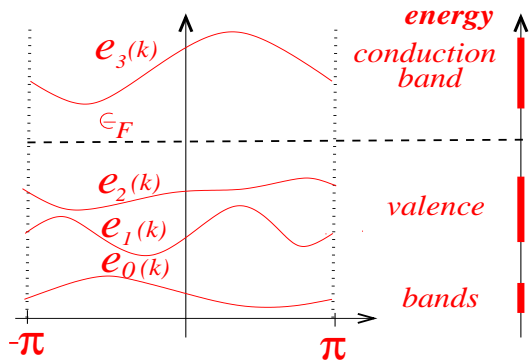}\\
\end{center}
\vskip -0.1cm
\caption{Energy bands ($1d$ cut) and the spectrum of $\,H_{_\CC}$}
\label{fig:energy_bands}
\end{figure}

Time evolution of the crystalline system is governed by a 
Hamiltonian $\,H_{_\CC}\,$ 
\qq
(H_{_\CC}\psi)(x)\,=\,\sum\limits_{y\in\CC}h(x,y)\,\psi(y)
\label{Hlattice}
\qqq
where $\,h(x,y)\in{\rm End}(\mathbb V)\,$ are assumed to vanish for $\,|y-x|>r\,$
for some fixed range $\,r>0$. We also assume that\footnote{This eliminates
external magnetic field so e.g. the Harper-Hofstadter model.} 
$\,h(x,y)=h(x+a,y+a)\,$ for $\,a\in\Gamma\,$ so that $\,[H_{_\CC},U(a)]=0$. 
Under isomorphism (\ref{FTr}), Hamiltonian $\,H_{_\CC}\,$ becomes a
family of Hermitian $\,N\times N\,$ matrices $\,\hat H(k)\,$ analytically  
depending on $\,k\in\BZ$. Diagonalization of $\,\hat H(k)\,$ leads to
energy bands $\,e_n(k)\,$ (that generically avoid crossing) such 
that 
\qq 
\hat H(k)\,\hat\psi_n(k)\,=\,e_n(k)\,\hat\psi_n(k)\,.
\qqq
$E\in{spec}(H_{_\CC})\,$ if and only if 
$\,E=e_n(k)\,$ for some $\,n\,$ and $\,k$,
\,see Fig.\,\ref{fig:energy_bands}.
We shall be interested in insulators where Hamiltonian $\,H_{_\CC}\,$ 
has a spectrum gap around the Fermi energy\footnote{This is 
the energy such that the ground state of the second-quantized 
systems has all states with energy $<\epsilon_F\,$ 
occupied and all states with energy $>\epsilon_F\,$ empty.} 
$\,\epsilon_F$. The bands with energies smaller than $\,\epsilon_F\,$ 
are called valence bands and the ones with energies bigger 
than $\,\epsilon_F\,$ are termed conduction bands. Below, we 
shall always assume that the Fermi energy $\,\epsilon_F=0$. This
may be achieved by subtracting non-zero $\,\epsilon_F\,$ from
Hamiltonian $\,H_{_\CC}$.

\subsection{Chern insulators and their homotopic invariant}
\label{subsec:Chern_ins}

\noindent Consider a crystalline insulator in two dimensions.
The simplest topological invariant of $\,2d\,$ insulators is
the $1^{\rm st}$ Chern number
\qq
c_1\,=\,\frac{_\ii}{^{2\pi}}\int_\BZ\,\tr\,P(k)\,\big(dP(k)\big)^{\wedge 2}\,,
\label{Chern}
\qqq
where $\,P(k)\,$ are the spectral projections of the Bloch
Hamiltonians $\,\hat H(k)\,$ on the negative eigenvalues (i.e. on the valence 
band energies). Geometrically, $\,c_1\,$ is the $1^{\rm st}$ Chern number
of the valence band sub-bundle $\,\CE\,$ of the trivial bundle 
$\,\BZ\times\mathbb C^N\,$ over the Brillouin torus whose fibers are 
spanned by the valence band eigenstates of $\,\hat H(k)$. The 
$1^{\rm st}$ Chern number is equal to the integral of the trace of 
the curvature of the Berry connection $\,\nabla^B\,$ on the sub-bundle 
$\,\CE\,$ divided by $\,2\pi$. \,It is an integer that is constant 
under continuous deformations of Hamiltonian $\,H_{_\CC}\,$ which do not 
close the spectral zero-energy gap.
\vskip 0.2cm
 
\begin{figure}[h]
\leavevmode
\begin{center}
\includegraphics*[width=5cm,height=8cm,angle=-90]{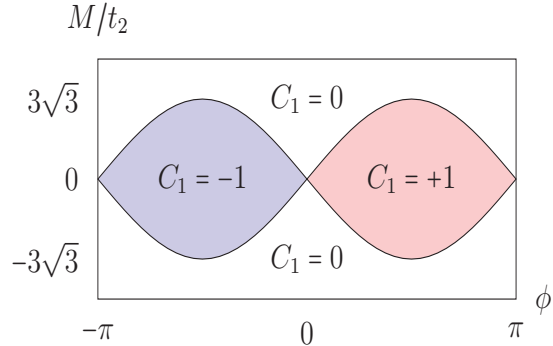}\\
\end{center}
\vskip -0.2cm
\caption{Phase diagram of the Haldane model (from \cite{CF})}
\label{fig:Chern_num}
\end{figure}

A simple example is provided by Hamiltonians 
\qq
\hat H(k)\,=\,\vec{d}(k)\cdot\vec{\sigma}
\label{H(k)}
\qqq
for $\,N=2$, \,where $\,\vec{\sigma}=(\sigma_x,\sigma_y,\sigma_z)\,$
is the vector of Pauli matrices. Spectrum of $\,\hat H(k)\,$
is composed of $\,\pm|\vec{d}(k)|$. \,In particular, for $\,|\vec{d}(k)|\,$
not vanishing on $\,\BZ$, \,the model describes an insulator. The
$1^{\rm st}$ Chern number is in this case equal to the degree of the
map 
\qq
\BZ\ni k\,\longmapsto\,\frac{\vec{d}(k)}{|\vec{d}(k)|}\in S^2\,.
\qqq
One of the simplest models with nontrivial $\,c_1\,$ was
proposed by Haldane \cite{Hald}. It lives on the hexagonal
crystal of \,Fig.\,\ref{fig:crystal} and has the tight binding Hamiltonian
\qq
H_{_\CC}\,=\,t\sum\limits_{n.n.}|x\rangle\langle y|\pm\ii\,t_2\sin(\varphi)
\sum\limits_{n.n.n}|x\rangle\langle y|+M\sum\limits_{x\in A}|x\rangle\langle x|
-M\sum\limits_{x\in B}|x\rangle\langle x|\,,
\qqq
where $\,n.n.\,$ denotes the nearest neighbors 
and $\,n.n.n.\,$ the next nearest neighbors $\,(x,y)\in\CC^2$,
\,and $\,A\,$ and $\,B\,$ denote two orbits of $\,\Gamma\,$ in
$\,\CC$. The signs in front of the $\,n.n.n\,$ terms are chosen as plus
if $\,x\in A\,$ and $\,y=x+a_1$, $\,y=x-a_1+a_2$, \,and $\,y=x-a_2$,
\,and if $\,x\in B\,$ and $\,y=x-a_1$, $\,y=x+a_1-a_2$, \,and $\,y=x+a_2$.
In the other half of cases the sign is chosen as minus. In the Fourier
transformed picture, $\,H_{_\CC}\,$ becomes a family of Hermitian matrices
(\ref{H(k)}) with
\qq
&d_x=t\big(1+\cos(k\cdot a_1)+\cos(k\cdot k_2)\big)\,,\qquad
d_y=t\big(\sin(k\cdot a_1)+\sin(k\cdot a_2)\big)\,,&\\
&d_z=M+2t_2\sin(\varphi)\big(\sin(k\cdot a_1)-\sin(k\cdot a_1-k\cdot a_2)
-\sin(k\cdot a_2)\big)\,.&
\qqq 
The phase diagram of $\,1^{\rm st}$ Chern numbers is given in 
Fig.\,\ref{fig:Chern_num}.
\vskip -0.6cm  
\noindent Numbers $\,c_1\,$ are obtained by counting with signs the points
in $\,\BZ\,$ where $\,\vec{d}(k)\,$ lies along the positive $\,z$-axis.
Such count may be viewed as a localization of the integral formula (\ref{Chern}).
\vskip 0.2cm

The $1^{\rm st}$ Chern number may be also obtained from a $3d$ integral.
Let 
\qq
\Phi(t,k)=\ee^{2\pi\ii\,tP(k)}=\ee^{2\pi\ii t}P(k)+I-P(k)\,.
\label{mapPhi}
\qqq
Note that $\,\Phi(t,k)\,$ are unitary matrices and that 
$\,\Phi(0,k)=I=\Phi(1,k)\,$ so that $\,\Phi\,$ may be viewed as a map
from a 2-sphere $\,S^2\,$ to the group $\,U(N)\,$ for which one may
consider the homotopic invariant 
\qq
{\rm deg}(\Phi)\,\equiv\,\frac{_1}{^{2\pi}}\hspace{-0.2cm}
\int\limits_{[0,1]\times\BZ}\hspace{-0.2cm}\Phi^*H
\ \,\in\,\mathbb Z\,,
\label{deg}
\qqq
where $\,H\,$ is the same bi-invariant 3-form on $\,U(N)\,$ as before
(not to be confused with the crystalline Hamiltonian $\,H_{_\CC}$).
\,A small calculation performing the $\,t$-integral shows that
\qq
{\rm deg}(\Phi)\,=\,\frac{_\ii}{^{2\pi}}
\int_\BZ\,\tr\,P(k)\,\big(dP(k)\big)^{\wedge 2}\,=\ c_1\,.
\label{deg1}
\qqq
\vskip 0.3cm

\noindent{\bf Remark.} \,Formula (\ref{deg1}) is a realization 
of the Bott isomorphism $\,\widetilde K^0(\BZ)\cong \widetilde K^1(S\BZ)\,$ 
between the $K$-theory groups, where $\,SM\,$ is the suspension 
of space $\,M$.

\subsection{Time reversal symmetry and the $2d$ Kane-Mele invariant}
\label{subsec:2d_KM_ind}

\begin{figure}[h]
\vskip -0.1cm
\leavevmode
\begin{center}
\includegraphics*[width=5.5cm,height=4.6cm,angle=0]{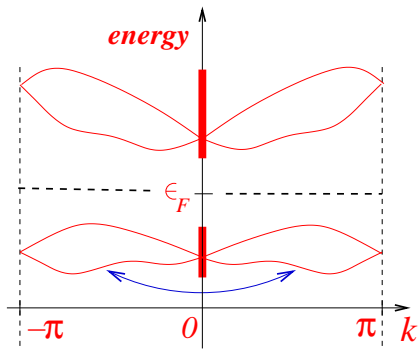}\\
\end{center}
\caption{Time reversal symmetric energy bands for $\,\theta^2=-I$} 
\label{fig:Kramers}
\end{figure}

\noindent Consider now the time reversal anti-unitary operator $\,\theta\,$ 
acting on the space of internal degrees of freedom $\,\mathbb V\,$ such 
that $\,\theta^2=\pm I$. It acts point-wise on the wave
functions $\,\psi(x)\in L^2(\CC,\mathbb V)\,$ and becomes under 
the Fourier transform (\ref{FTr}) the operator
\qq
\hat\psi\,\longmapsto \theta\,\hat\psi\circ\vartheta\,,
\qqq
where $\,\theta\,$ is viewed now as the anti-unitary operator
in $\,L^2(F,\mathbb V)\cong\mathbb C^N$ for $\,F\,$ a fixed unit cell 
in $\,\CC\,$ and where $\,\vartheta:\BZ\rightarrow\BZ\,$ is induced by 
$\,k\mapsto-k$. \,For time-reversal symmetric crystalline Hamiltonians 
$\,\theta H_{_\CC}\theta^{-1}=H_{_\CC}$,
\qq
\theta\,\hat H(k)\,\theta^{-1}\,=\,\hat H(-k)\,
\label{TRH}. 
\qqq
If $\,\hat\psi_n(k)\,$ is an eigenvector of $\,\hat H(k)\,$ then 
$\,\theta\hat\psi_n(k)\,$ is an eigenvector of $\,\hat H(-k)\,$
with the same eigenvalue. As the result, the band spectrum is symmetric 
under $\,k\mapsto -k+b\,$ for each $\,b\,$ in the reciprocal lattice 
$\,\Gamma^r$. Finally, if $\,\theta^2=-I\,$ then vectors 
$\,\theta\hat\psi_n(k)\,$ and $\,\hat\psi_n(k)\,$ are orthogonal. Indeed,
\qq
\langle\hat\psi|\theta\hat\psi\rangle=\langle\theta^2\hat\psi|
\theta\hat\psi\rangle=-\langle\hat\psi|\theta\hat\psi\rangle=0\,.
\qqq
Such pairs of eigenvectors vectors in $\,\mathbb C^N\,$ are called Kramers 
pairs. In particular, for time-reversal symmetric momenta $\,k=-k+b\,$ 
(the so called TRIM), the spectrum of $\,\hat H(k)\,$ has even degeneration 
and the typical picture of bands is as in Fig.\,\ref{fig:Kramers}. In 
particular, the dimension of the range of the projectors $\,P(k)\,$ on 
the valence band states is even.

\noindent Symmetry (\ref{TRH}) implies that
\qq
\theta\,P(k)\,\theta^{-1}\,=\,P(-k)
\label{P(k)TRS}
\qqq
and, consequently, that the corresponding Chern number vanishes. Indeed,
\qq
c_1\,=\,\frac{\ii}{2\pi}\int_\BZ\,\tr\,\theta\,P(-k)\,\big(dP(-k)\big)^{\wedge 2}
\theta^{-1}\,=\,\frac{\ii}{2\pi}\int_\BZ\,\overline{\tr\,P(-k)\,
\big(dP(-k)\big)^{\wedge 2}}\,=\,-c_1\,=\,0\,,
\qqq
since the 2-form $\,\tr\,P(dP)^{\wedge 2}\,$ is purely
imaginary. The vanishing of the $1^{\rm st}$ Chern number implies that
the vector bundle $\,\CE\,$ over $\,\BZ\,$ formed by the valence band states 
is topologically trivial. In \cite{KM}, Kane and Mele observed, however,
that, for $\,\theta^2=-I$, vector bundle $\,\CE\,$ (of even rank $\,m$) 
\,may still exhibit nontrivial topological properties when considered 
together with the action of $\,\theta\,$ mapping anti-unitarily fibers 
$\,\CE_k\,$ to $\,\CE_{-k}$. The nontrivial topological property is captured 
by an obstruction to the existence of a global frame of sections 
$\,(\hat\psi_1(k),\dots,\hat\psi_{m}(k))\,$ of $\,\CE\,$ (not necessarily 
composed of eigenfunctions of $\,\hat H(k)$) \,such that 
$\,\hat\psi_{2i}(k)=\theta\hat\psi_{2i-1}(-k)$, \,i.e. formed by 
globally defined Kramers pairs. Kane and Mele showed that the obstruction
is given by a $\,\mathbb Z_2$-valued index, that we shall denote by $\,\KM\,$ 
and describe in the form established in \cite{FK}.
\vskip 0.2cm

One chooses an arbitrary global frame $\,(\hat\psi_1(k),\dots,
\hat\psi_{m}(k))\,$ trivializing bundle $\,\CE\,$ and defines the 
``sewing'' matrix
\qq
w_{ij}(k)\,=\,\langle \hat\psi_i(-k)|\theta\hat\psi_j(k)\rangle\,.
\label{wk}
\qqq
$w(k)=(w_{ij}(k))\,$ is a unitary $\,m\times m$ matrix that satisfies
$\,w(-k)=-w(k)^T$. \,It follows that $\,\det(w(k))=\det(w(-k))\,$ so that 
the determinant of $\,w(k)\,$ does not wind around the cycles of $\,\BZ\,$ and, 
consequently, it has a globally defined square root determined up to an overall 
sign. Besides, at four TRIM $\,k\in\BZ$, $\,w(k)\,$ is an antisymmetric matrix. 
The Kane-Mele index $\,\KM\,$ is given by:
\qq
(-1)^{\KM}\,=\,\prod\limits_{{\rm TRIM}\ k}\frac{\sqrt{\det(w(k))}}{{\rm pf}(w(k))}
\,=\,\ee^{\ii S_\WZ(w)}\,,
\label{KaneMele}
\qqq
where $\,{\rm pf}\,$ denotes the Pfaffian of an antisymmetric matrix.
The first equality is the expression of ref.\,\cite{FK} whereas the
second equality seems to be new and will be discussed elsewhere.
It was observed already in the original paper \cite{KM} that $\,\KM\,$
lives in the Real $K$-theory group of the $2d$-torus $\,\BZ\,$ equipped with 
involution $\,\vartheta\,$ \cite{Atiyah}, \,namely in the reduced group 
$\,\widetilde{K\hspace{-0.03cm}R}^{-4}(\BZ)\cong\mathbb Z_2$. 
\,Index $\,\KM\,$ is invariant under deformations of the crystalline system 
preserving the time reversal symmetry and the gap. 
\vskip 0.2cm

Kane and Mele discussed in \cite{KM} a concrete model with a 
nontrivial value of $\,\KM$. \,The model was obtained by coupling two 
copies of the Haldane model corresponding to two components of 
spin$\,\,=\frac{1}{2}$ by an additional interaction between two spin 
components (mimicking the spin-orbit interaction). A somewhat different 
version of a model with $\,\KM\not=0\,$ was proposed in \cite{BHZ}. 
The nontrivial topological phases with $\,\KM=-1\,$ were soon
realized experimentally \cite{KWBRBMQZ}. Mathematical proofs that $\,\KM\,$
realizes the obstruction to the global choice of a frame of vector bundle
$\,\CE\,$ composed of Kramers pairs were given in \cite{DNG,FMP}.

\subsection{Kane-Mele invariant as the square root of \,a WZ amplitude}
\label{subsec:KM_sqrt}

\noindent In ref.\,\cite{CDFGT}, \,we showed that the invariant $\,\KM\,$ may be 
represented as the square root of the WZ amplitude of the field
\qq
\BZ\,\ni\,k\ \longmapsto\ \phi(k)\,=\,I-2P(k)\,\in\,U(N)
\qqq
that satisfies the equivariance condition (\ref{thetaphi}), \,see
(\ref{P(k)TRS}). Our formula (\ref{WZKM}) for $\,\KM\,$ seems to provide a 
realization unknown in mathematical literature of the Bott 
isomorphism in Real $K$-theory $\,\widetilde{K\hspace{-0.03cm}R}^{-4}(\BZ)
\cong\widetilde{K\hspace{-0.03cm}R}^{-3}(S\BZ)$.
The proof given in \cite{CDFGT} that $\,\sqrt{\ee^{\ii S_\WZ(\phi)}}\,$ is equal 
to $\,(-1)^\KM\,$ was based on the localization at TRIM of the nonlocal 
formula (\ref{sqrtWZ}) for the square root of the WZ amplitude of $\,\phi$.  
\,Although rather laborious, it involved interesting elements such as 
a new boundary gauge anomaly for WZ amplitudes. Here, we shall show that 
definition (\ref{sqrthatS}) and the local expression (\ref{sqrtHol2}) provide 
a more direct way to establish relation (\ref{WZKM}) resembling the derivation 
of the Kane-Mele invariant in \cite{FK}.  
\vskip 0.2cm

In order to calculate 
$\,\sqrt{\ee^{\ii S_\WZ(\phi)}}=\sqrt{Hol_{\widehat\CG}(\widehat\phi)}\,$ 
using (\ref{sqrtHol2}), let us start by choosing as (the closure of) 
a fundamental domain $\,F\subset \BZ\,$ the ``effective Brillouin 
zone'' 
\qq
\BZ_+\,=\,\big\{k\in\BZ\,\big|\,0\leq k_1\leq\pi\big\}\,,
\qqq
as on the left hand side of \,Fig.\,\ref{fig:triangul}.
Let $\,\CG=(Y,B,\CL,t)\,$ be the basic gerbe on $\,U(N)\,$
constructed in Sec.\,\ref{subsec:basic_U(N)}. Consider map 
$\,s:\BZ_+\rightarrow Y\,$ defined by 
\qq
s(k)=(-\frac{_{3}}{^2}\pi,\phi(k))
\qqq
which is well defined because $\,spec(\phi(k))\subset\{\pm1\}$.
It satisfies the relation $\,\pi\circ s=\phi|_F$. \,In the calculation of
$\,Hol_\CG(\phi|_F)\,$ as given by (\ref{HolCG}) we shall use $\,s_c=s|_c\,$
for all triangles of a triangulation of $\,\BZ_+$.
\,Similarly, we shall set $\,s_b=s|_b\,$ for all edges $\,b\subset F\,$
except those in $\,\ell\,$ for $\,\ell\subset\partial F\,$ 
chosen as on the left hand side of \,Fig.\,\ref{fig:triangul}.
For those, we shall take $\,s_b=s_\ell|_b$, \,where
$\,s_\ell:\ell\rightarrow Y\,$ is given by 
\qq
s_\ell(k)\,=\,(-\frac{_1}{^2}\pi,\phi(k))\in Y\,.
\qqq
This will ensure relation (\ref{sbsb})
for $\,b\subset\ell$. \,Finally, for $\,v\in\partial\ell$, \,we shall
set $\,s_v=s_\ell(v)=(-\frac{_1}{^2}\pi,\phi(v))\in Y$.
\,With those choices, the expression for the gerbe holonomy of $\,\phi|_F\,$  
simplifies to
\qq
Hol_\CG(\phi|_F)\,=\ \ee^{\ii\sum\limits_{c\subset F}\int_cs_c^*B}
\mathop{\otimes}\limits_{b\subset c\subset F}hol_\CL(s_c|_b,s_b)
\,=\ \ee^{\ii\int_{\BZ_+}s^*B}\,hol_\CL(s|_\ell,s_\ell)\ \in\,\mathop{\otimes}
\limits_{v\in\partial\ell}\CL_{s(v),s_v}\,.
\label{forHolF}
\qqq
A direct calculation using the relation $\,H^{\rm eff}_{-\frac{3}{2}\pi}(\phi(k))
=-\pi P(k)$ and the identity $\,(PdP)^2=0\,$ shows that
\qq
s^*B\,=\,
-\frac{_\ii}{^2}\,\tr\,P(dP)^{\wedge 2}\,. 
\qqq
Let $\,(\widehat\psi_i(k))_{i=1}^m\,$ be a global orthonormal frame of
the valence vector bundle $\,\CE\,$ whose existence is guaranteed
by the vanishing of the $1^{\rm st}$ Chern number of $\,\CE$. \,One has
the relations
\qq
P(k)=\sum\limits_{i=1}^m|\widehat\psi_i(k)\rangle\langle\widehat\psi_i(k)|\,,
\qquad A^B(k)=\ii\sum\limits_{i=1}^m\langle\widehat\psi_i(k)|d\widehat
\psi_i(k)\rangle\,,
\qqq
where $\,A^B\,$ is the Berry connection 1-form corresponding to the
curvature 2-form $\,F^B=dA^B=\ii\,\tr\,P(dP)^{\wedge2}$. \,Hence,
by the Stokes formula,
\qq
\int_Fs^*B\,=\,-\frac{_1}{^2}\int_FdA^B\,=\,
-\frac{_1}{^2}\int_{\partial F}A^B\,=\,
-\frac{_1}{^2}\int_{\ell}(A^B-\vartheta^*A^B)\,,
\qqq
where we used the relation $\partial F=\ell\cup\vartheta(\ell)$. \,Now
\qq
(\vartheta^*A^B)(k)\,=\,\ii\sum\limits_{i=1}^m\langle\widehat\psi_i(-k)|
d\widehat\psi_i(-k)\rangle\,=\,
\ii\sum\limits_{i=1}^m\langle\theta d\widehat\psi_i(-k)|\theta
\widehat\psi_i(-k)\rangle\,.
\qqq
Substituting to the right hand side the relation
$\,\theta\widehat\psi_i(-k)=
-\sum_{j}w_{ij}(k)\,\widehat\psi_j(k)$,
\,where $\,w_{ij}(k)=-w_{ji}(-k)\,$ in the sewing matrix of (\ref{wk}),
\,we infer that
\qq
\vartheta^*A^B
\,=\,-\ii\,\tr\,w^{-1}dw\,-\,A^B
\qqq
and, consequently, that
\qq
\int_Fs^*B=-\int_\ell A^B\,-\,\frac{_\ii}{^2}\int_\ell
\tr\,w^{-1}dw\,.
\label{intsB}
\qqq
On the other hand, the map $\,(s|_\ell,s_\ell)\,$ from $\,\ell\,$ to 
$\,Y^{[2]}\,$ is given by $\,(s(k),s_\ell(k))=(-\frac{3}{2}\pi,
-\frac{1}{2}\pi,\phi(k))$. \,Since $\,P_{-\frac{3}{2}\pi,
-\frac{1}{2}\pi}(\phi(k))=P(k)$, it follows that
$\,\CL_{s(k),s_\ell(k)}\,=\,\wedge^{\rm max}P(k)\,\mathbb C^N\,$
and the parallel lift of $\,(s|_\ell,s_\ell)\,$ to $\,\CL\,$ is given by
\qq
\ee^{\ii\int\limits_0^{k_2}\big(A^B(a,k'_2)+A(-\frac{_3}{^2}\pi,
-\frac{_1}{^2}\pi,\phi(a,k'_2))\big)}\,\widehat\psi_1(a,k_2)
\wedge\cdots\wedge\widehat\psi_m(a,k_2)\,.
\label{partr}
\qqq
for $\,k_2\in[0,\pi]\,$ and $\,a=0\,$ or $\,a=\pi\,$ corresponding to 
one of the two connected components of $\,\ell$. \,But 
\qq
&A(-\frac{_3}{^2}\pi,-\frac{_1}{^2}\pi,\phi(a,k'_2))\,=\,
-\frac{_1}{^4}\,\tr\big(H^{\rm eff}_{-\frac{2}{2}\pi}(\phi(a,k_2'))\,dP(a,k'_2)
\big)&\cr
&=\,\frac{_\pi}{^4}\,\tr\big(P(a,k_2'))\,dP(a,k'_2)\big)\,=\,
\frac{_\pi}{^8}\,\tr\big(dP(a,k'_2)\big)\,=\,0\,.&
\qqq
As the result, one obtains from (\ref{partr}) the following expression for 
the line bundle $\,\CL\,$ holonomy term:
\qq
hol_\CL(s|_\ell,s_\ell)\,=\ 
\ee^{\ii\int_\ell A^B}\mathop{\otimes}\limits_{v\in\partial\ell}
\big(\widehat\psi_1(v)\wedge\cdots\wedge\widehat\psi_m(v)\big)^{\pm1}.
\label{holss}
\qqq
From (\ref{forHolF}), (\ref{intsB}) and (\ref{holss}), one then infers that
\qq
Hol_\CG(\phi|_F)\,=\,\ee^{\frac{_{1}}{^2}\int_\ell{\rm tr}\,w^{-1}dw}
\mathop{\otimes}\limits_{v\in\partial\ell}
\big(\widehat\psi_1(v)\wedge\cdots\wedge\widehat\psi_m(v)\big)^{\pm1}\,\in\,
\mathop{\otimes}\limits_{v\in\partial\ell}\CL_{\Theta_Y(s_v),s_v}^{\pm1}\,,
\qqq

Now, the triviality of line bundle $\,\CN$, \,for the 1-isomorphism 
$\,\eta=(\CN,\nu):\CG\rightarrow\Theta^*\CG\,$ constructed in
Sec.\,\ref{subsec:Theta_equiv}, implies that
\qq
hol_\CN(\phi|_\ell)\,=\,1\,\in\mathop{\otimes}\limits_{v\in\partial\ell}
\CN_{s_v}\,.
\qqq
Hence
\qq
\hspace*{-0.4cm}Hol_\CG(\phi|_F)\,\otimes\,hol_\CN(\phi|_\ell)&=&
\,\ee^{-\frac{_{\ii}}{^2}\int\limits_\ell{\rm tr}\,w^{-1}dw}
\mathop{\otimes}\limits_{v\in\partial\ell}
\big(\widehat\psi_1(v)\wedge\cdots\wedge\widehat\psi_m(v)\big)^{\pm1}\cr
&=&\,\ee^{\frac{_{1}}{^2}\int\limits_\ell{\rm tr}\,w^{-1}dw}
\,\big(\widehat\psi_1(\pi,0)\wedge\cdots\wedge
\widehat\psi_m(\pi,0)\big)^{-1}\otimes\big(\widehat\psi_1(\pi,\pi)\wedge\cdots\wedge\widehat\psi_m(\pi,\pi)
\big)\cr
&&\hspace{1.65cm}\otimes\,
\big(\widehat\psi_1(0,\pi)\wedge\cdots\wedge\widehat\psi_m(0,\pi)\big)^{-1}\otimes
\big(\widehat\psi_1(0,0)\wedge\cdots\wedge\widehat\psi_m(0,0)\big)
\label{posred}\\ 
&&\hspace{1.1cm}\in\ 
\CN'^{-1}_{(-\frac{1}{2}\pi,\phi(\pi,0))}\otimes\CN'_{(-\frac{1}{2}\pi,\phi(\pi,\pi))}
\otimes\CN'^{-1}_{(-\frac{1}{2}\pi,\phi(0,\pi))}\otimes
\CN'_{(-\frac{1}{2}\pi,\phi(0,0))}\,.
\nonumber
\qqq
Using the sections $\,\sqrt{S'_\pm}\,$ of the line bundle $\,N'\,$ determined
by assignment (\ref{S'pm1/2}), 
\,and noting that when we lift $\,\phi(k)=I-2P(k)\,$ to a map 
$\,\widehat\phi(k)\,$ with values in 
$\,\widehat U(N)\,$ then at all TRIM $\,\widehat\phi(k)\,$ is in the same
component of $\,\widehat U(N)'\,$ because $\,\det(\phi(k)=1\,$ for all $\,k$, 
\,the last expression may be rewritten as
\qq
&&Hol_\CG(\phi|_F)\,\otimes\,hol_\CN(\phi|_\ell)\ =\ 
\ee^{\frac{_{1}}{^2}\int\limits_\ell{\rm tr}\,w^{-1}dw}
\,\frac{{\rm pf}(w(\pi,0))\,{\rm pf}(w(0,\pi))}
{{\rm pf}(w(\pi,\pi))\,{\rm pf}(w(0,0))}\cr\cr
&&\hspace{0.5cm}\times\,\sqrt{S'_\pm}(\phi(\pi,0))^{-1}
\otimes\sqrt{S'_\pm}(\phi(\pi,\pi))\otimes\sqrt{S'_\pm}(\phi(0,\pi))^{-1}
\otimes\sqrt{S'_\pm}(\phi(0,0))\cr\cr
&&\equiv\ \sqrt{\ee^{\ii S_{\WZ}(\phi)}}\ \sqrt{S'_\pm}(\phi(\pi,0))^{-1}
\otimes\sqrt{S'_\pm}(\phi(\pi,\pi))\otimes\sqrt{S'_\pm}(\phi(0,\pi))^{-1}
\otimes\sqrt{S'_\pm}(\phi(0,0))\quad
\qqq
with the same sign in all $\,\sqrt{S'_\pm}$. \,This gives 
\qq
\sqrt{\ee^{\ii S_{\WZ}(\phi)}}&=&\,\ee^{\frac{_{1}}{^2}
\int_\ell{\rm tr}\,w^{-1}dw}
\,\frac{{\rm pf}(w(\pi,0))\,{\rm pf}(w(0,\pi))}
{{\rm pf}(w(\pi,\pi))\,{\rm pf}(w(0,0))}\cr\cr
&=&\,\frac{\sqrt{\det(w(\pi,\pi))}\,\sqrt{\det(w(0,0))}}
{\sqrt{\det(w(\pi,0))}\,\sqrt{\det(w(0,\pi))}}
\,\frac{{\rm pf}(w(\pi,0))\,{\rm pf}(w(0,\pi))}
{{\rm pf}(w(\pi,\pi))\,{\rm pf}(w(0,0))}\,,
\qqq
which is equivalent to expression (\ref{KaneMele}) for $\,(-1)^\KM\,$ given in
\cite{FK} (recall  that $\,\sqrt{\det(w(k))}\,$ is defined on $\,\BZ\,$
modulo a global sign).
\vskip 0.2cm

Let us mention another possible representation of 
$\,(-1)^\KM$. \,To this end we observe that field $\,\Phi\,$ of \,(\ref{mapPhi}) 
may be considered as a map
\qq
R\,\equiv\,\mathbb R/\mathbb Z\times\BZ\,\ni\,(t,k)\ 
\longmapsto\ \Phi(t,k)\,\in\,U(N)
\qqq
satisfying equivariance condition (\ref{equivPhi})
for the orientation reversing involution $\,\rho(t,k)=(-t,-k)$.
\,Indeed,
\qq
\Phi(-t,-k)=\ee^{-2\pi\ii tP(-k)}=\ee^{-2\pi\ii t\,\theta P(k)\theta}=
\theta\,\ee^{2\pi\ii t P(k)}\theta^{-1}=\theta\,\Phi(t,k)\theta^{-1}\,.
\qqq
Let us consider the $3d$ index defined in Sec.\,\ref{subsec:3d_ind},
\qq
\CK(\Phi)\,=\,\frac{\ee^{\frac{\ii}{2}\int_{F_R}\Phi^*H}}
{\sqrt{\ee^{\ii S_\WZ(\Phi|_{\partial F_R})}}}\,,
\label{3dUP}
\qqq
for the fundamental domain $\,F_R=[0,\frac{1}{2}]\times\BZ\,$
of involution $\,\rho$. \,Now, field $\,\Phi\,$ also satisfies the
relation
\qq
\Phi(t,-k)=\theta\,\Phi(t,k)^{-1}\theta^{-1}\,=\Theta\circ\Phi^{-1}(t,k)
\qqq
which implies that $\,(\Phi^*H)(t,-k)=((\Phi^{-1})^*H)(t,k)=-(\Phi^*H)(t,k)\,$
and, consequently, that the numerator in (\ref{3dUP}) vanishes. Hence
\qq
\CK(\Phi)\,=\,\frac{1}{\sqrt{\ee^{\ii S_\WZ(\Phi|_{\partial F_R})}}}
\,=\,\sqrt{\ee^{\ii S_\WZ(\Phi|_{\partial F_R})}}\,=\,(-1)^\KM,
\label{KMWZ}
\qqq
where the last but one equality follows from the fact that $\,\CK(\Phi)=\pm1$.
The gain from the representation (\ref{KMWZ}) of the Kane-Mele invariant is 
that index $\,\CK(\Phi)\,$ may be computed by using other choices
of the fundamental domain $\,F_R$, \,e.g. by taking $\,F_R=(\mathbb R/\mathbb Z)
\times\BZ_+$. \,The corresponding formula for $\,\KM\,$ is close, at least 
in the spirit, to the $\,\mathbb Z_2$-valued index defined in \cite{MB}, 
which was not shown previously to be equal to the Kane-Mele index.

\subsection{$3d$ \,Kane-Mele invariants}
\label{subsec:3d_KM_ind}
 
\begin{figure}[ht]
\leavevmode
\begin{center}
\includegraphics*[width=5.4cm,height=4.4cm,angle=0]{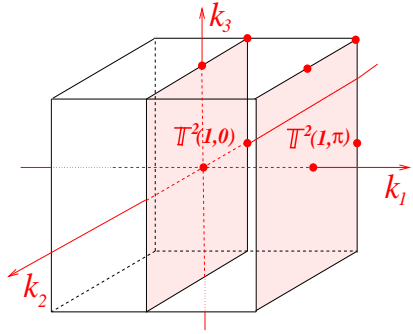}\\
\end{center}
\caption{Two of six $\,\vartheta\,$ invariant sub-tori in $3d$ Brillouin torus}
\label{fig:torus3d}
\end{figure}

\noindent In ref.\,\cite{FKM}, \,Fu, Kane and Mele extended their index to $3d$ 
time-reversal insulators. In its strong form which is related to a $K$-theory 
class in $\,\widetilde{K\hspace{-0.03cm}R}^{-4}(\BZ)\,$ pulled back from 
$\,\widetilde{K\hspace{-0.03cm}R}^{-4}(S^3)\cong\mathbb Z^2$, \,where here 
$\,\BZ\,$ is the $3d$ Brillouin torus $\,\cong\mathbb R^3/(2\pi\mathbb Z^3)\,$ 
with the orientation-reversing involution $\,\rho\,$ induced by 
$\,k\mapsto-k\,$ in $\,\mathbb R^3$. 
The strong Kane-Mele index $\,\KM^s\,$ is still defined by the first 
equality in (\ref{KaneMele}) 
but the product is now over $\,8\,$ TRIM is $\,\BZ$. 
\,On the other hand, there are $\,6\,$ weak $2d$ indices $\,\KM({i,a})\,$ for
$\,i=1,2,3\,$ and $\,a=0,\pi\,$ obtained by 
restricting the products over TRIM to one of six $\,\rho$-invariant $2d$ 
sub-tori $\,\mathbb T^2({i,a})\subset\BZ$, \,see Fig.\,\ref{fig:torus3d}.
Clearly, the strong index $\,\KM^s\,$ is equal to the sum or difference 
(mod\ 2) of two weak indices corresponding to each pair of $2d$ 
$\,\rho$-invariant sub-tori perpendicular to a coordinate axis:
\qq
\KM^s\,=\,\KM({i,\pi})-\KM({i,0})
\label{KMsKMKM}
\qqq
for each $\,i=1,2,3$.
\vskip 0.3cm

For the map
\qq
\BZ\ni k\ \longmapsto\ \Phi(k)=I-2P(k)\in U(N)\,,
\qqq
where $\,P(k)\,$ are the spectral projectors on the valance band states
of the $3d$ time reversal invariant insulator, the equivariance
\qq
\Phi\circ\rho\,=\,\Theta\circ\Phi
\qqq
is assured by the relation $\,P(-k)=\theta P(k)\theta^{-1}\,$
following from the time-reversal invariance of the Hamiltonian. We may then
consider the $3d$ index $\,\CK(\Phi)\,$ with values $\,\pm1\,$ defined in
Sec.\,\ref{subsec:3d_ind}. Taking for the the fundamental domain 
$\,F_R\subset\BZ\,$ the $3d$ effective Brillouin zone $\,\BZ_+\,$
corresponding to $\,k_1\in[0,\pi]$, \,we obtain:
\qq
\CK(\Phi)\,=\,\frac{\ee^{\frac{\ii}{2}\int_{\BZ_+}\Phi^*H}}
{\sqrt{\ee^{\ii S_\WZ(\Phi|_{\mathbb T^2({1,\pi})})}}\Big/
\sqrt{\ee^{\ii S_\WZ(\Phi|_{\mathbb T^2({1,0})})}}}\,,
\qqq
where $\,\mathbb T^2({1,a})\,$ for $\,a=0,\pi\,$ are two boundary
$2d$ sub-tori in $\,\partial\BZ_+$, \,see Fig.\,\ref{fig:torus3d}.
\,We have the following
\vskip 0.4cm

\noindent{\bf Lemma}. $\ \int_{\BZ_+}\hspace{-0.15cm}\Phi^*H\,=\,0\,$.
\vskip 0.3cm

\noindent{\bf Proof}.\ \,Consider the mapping
\qq
\widetilde{\BZ}_+\,\equiv\,[0,\frac{_1}{^2}]\times\BZ_+\,\ni\,(t,k)\ 
\longmapsto\ \widetilde\Phi(t,k)=
\ee^{2\pi\ii t P(k)}\,\in\,U(N)\,.
\qqq
Since $\,H\,$ is a closed 3-form, it follows from the Stokes Theorem that
\qq
\int_{\partial\widetilde{\BZ}_+}\widetilde\Phi^*H\,=\,0\,.
\label{reltbp}
\qqq
On the other hand,
\qq
\partial\widetilde{\BZ}_+\,=\,\{\frac{_1}{^2}\}\times\BZ_+\,-\,
\{0\}\times\BZ_+\,-\,[0,\frac{_1}{^2}]\times\mathbb T^2({1,\pi})\,+\,
\,[0,\frac{_1}{^2}]\times\mathbb T^2({1,0})\,, 
\qqq
and the contribution to the integral over $\,\partial\widetilde{\BZ}_+\,$
of the first piece is equal to $\,\int_{\BZ_+}\hspace{-0.15cm}\Phi^*H\,$
whereas the contributions of the other pieces vanish (on the
the second piece, $\,\widetilde\Phi=I\,$ and for the 
the third and the fourth piece, we use the same argument that
proved the vanishing of the numerator in (\ref{3dUP})). This
establishes relation (\ref{reltbp}).

\hspace{13cm}$\Box$
\vskip 0.2cm

\noindent Using the last lemma, we infer that
\qq
\CK(\Phi)\,=\,
\frac{\sqrt{\ee^{\ii S_\WZ(\Phi|_{\mathbb T^{1,0}})}}}
{\sqrt{\ee^{\ii S_\WZ(\Phi|_{\mathbb T^{1,\pi}})}}}\,,
\qqq
By (\ref{WZKM}), \,the square-roots of the $\,\WZ\,$ amplitudes
give the two weak $\,\KM\,$ indices and we finally obtain the
relation
\qq
\CK(\Phi)\,=\,(-1)^{\KM^s}
\label{KMsannoun} 
\qqq
announced in Sec.\,\ref{subsec:3d_ind}.
\vskip 0.1cm

The index $\,(-1)^{\KM^s}\,$  is equal to the Chern-Simons amplitude
of the connection with covariant derivative $\,\nabla S(k)=P(k)dS(k)\,$ 
on the (topologically trivial) valence vector bundle $\,\CE$, \,see \cite{FM}
and references therein. This can be easily proven directly from the second 
equality in (\ref{KaneMele}) and will be discussed elsewhere.

\subsection{Floquet theory of periodically driven crystalline systems}
\label{subsec:Floquet_th}

\noindent There exist an interesting possibility that one can induce 
nontrivial topological properties of materials by external forcing, for example
by irradiating a sample of the material with microwaves \cite{LRG}. Although
such a scenario for inducing topological phases has been realized up 
to now only in artificial systems made of arrays of coupled waveguides
or of cold atoms in optical lattices, the idea has inspired a considerable 
theoretical activity. 
\vskip 0.2cm

\begin{figure}[ht]
\leavevmode
\begin{center}
\hspace*{-1cm}
\includegraphics*[width=6cm,height=4.2cm,angle=0]{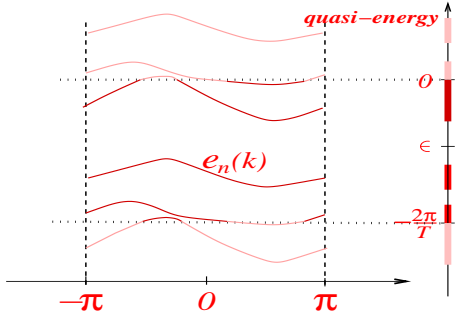}\\
\end{center}
\vskip -0.2cm
\caption{Quasi-energy bands}
\label{fig:qen_bands}
\end{figure}

Suppose that the Hamiltonian of a crystalline system acting in
space of states $\,\CH=L^2(\CC,\mathbb V)\,$ depends periodically on time, \,i.e.
\qq
H_{_\CC}(t+T)\,=\,H_{_\CC}(t)\,,
\qqq
for $\,H_{_\CC}(t)\,$ of the form (\ref{Hlattice}) with $\,h(t,x,y)=h(t+T,x,y)=
h(t,x+a,y+a)\,$ for $\,a\in\Gamma$. \,Such Hamiltonians
generate evolution operators $\,U_{_\CC}(t)\,$ satisfying
\qq
\ii\partial_tU_{_\CC}(t)\,=\,H_{_\CC}(t)\,U_{_\CC}(t)\,,\qquad U_{_\CC}(0)=I
\qqq
which are not periodic but satisfy the relation
\qq
U_{_\CC}(t+T)\,=\,U_{_\CC}(t)\,U_{_\CC}(T)\,.
\qqq 
In the Fourier picture, we obtain (analytic in $\,k$) families of
periodic in time Hermitian $N\times N$ matrices $\,\hat H(t,k)=\hat H(t+T,k)\,$ 
and of unitary $N\times N$ matrices $\,\hat U(t,k)\,$ satisfying 
the relation
\qq
\hat U(t+T,k)\,=\,\hat U(t,k)\,\hat U(T,k)\,.
\qqq
The main idea of the Floquet theory of periodically driven systems
is to replace the Bloch diagonalization of static Hamiltonians by
the diagonalization of the evolution operators over one
period of time. One just looks for the eigenvalues and eigenfunctions
such that
\qq
\hat U(T,k)\,\hat\psi_n(k)\,=\,\ee^{-\ii Te_n}\,\hat\psi_n(k)\,.
\qqq
We parameterized the eigenvalues of unitary matrices $\,\hat U(T,k)\,$
by ``quasi-energies'' $\,e_n(k)\,$ defined modulo $\,\frac{2\pi}{T}$.
They form a pattern of bands in $\,\BZ\times\mathbb R\,$ that repeats
itself with this period, \,see Fig.\,\ref{fig:qen_bands}.
Note that the time-dependent states 
$\,\hat\psi_n(t,k)=\hat U(t,k)\,\hat\psi_n(k)\,$ satisfy the relation
\qq
\ii\partial_t\hat\psi_n(t,k)\,=\,\hat H(t,k)\,\hat\psi_n(t,k)\,,\qquad
\hat\psi_n(t+T,k)\,=\,\ee^{-\ii Te_n(k)}\,\hat\psi_n(t,k)\,.
\qqq
They are called Floquet states.

\subsection{Topological invariants of gapped Floquet $2d$ and
$3d$ systems}
\label{subsec:inv_Floquet}

\noindent Suppose that there is a gap in the quasi-energy spectrum of 
$\,\hat U(T,k)\,$ around $\,\epsilon\,$ with $\,-\frac{2\pi}{T}<\epsilon<0\,$ 
for all $\,k\in\BZ$. \,Like in (\ref{Heff0}), we may then define  
the effective Hamiltonians
\qq
H^{\rm eff}_{\epsilon}(k)\,=\,\frac{_\ii}{^T}\,\ln_{-\epsilon T}\hat U(T,k)
\qqq
that depend analytically on $\,k\in\BZ\,$
and satisfy
\qq
\hat U(T,k)\,=\,\ee^{-\ii T H^{\rm eff}_\epsilon(k)}  
\qqq
(the definitions coincide with the ones of Sec.\,\ref{subsec:basic_U(N)} 
if the period $\,T\,$ is set to $\,1$). For two gap quasi-energies $\,
-\frac{2\pi}{T}<\epsilon_1\leq\epsilon_2<0$, 
\qq
H_{\epsilon_2}^{\rm eff}(k)-H_{\epsilon_1}^{\rm eff}(k)
=\frac{_{2\pi}}{^T}\,P_{\epsilon_1,\epsilon_2}(k)\,,
\label{2epsilons}
\qqq
where $\,P_{\epsilon_1,\epsilon_2}(k)\,$ is the spectral projection of $\,\hat 
U(T,k)\,$ on the part of the spectrum  in the sub-interval of the circle joining
$\,\ee^{-\ii T\epsilon_1}\,$ to $\,\ee^{-\ii T\epsilon_2}\,$ clockwise, 
\,see (\ref{tildeP}).
\vskip 0.2cm

Effective Hamiltonian may be used to define the periodized evolution operators
\qq
V_\epsilon(t,k)\,=\,\hat U(t,k)\,\ee^{\ii t H^{\rm eff}_\epsilon(k)}
\label{Vepsilon}
\qqq
that satisfy the relations
\qq
V_\epsilon(t+T,k)\,=\,V_\epsilon(t,k)\,,\qquad V_\epsilon(0,k)=I=V_\epsilon(T,k)\,.
\qqq
In \cite{RLBL}, the authors considered a topological invariant $\,W_\epsilon\,$
of the $2d$ gapped Floquet systems defined by
\qq
W_\epsilon\,=\,{\rm deg}(V_\epsilon)\,\equiv\,\frac{_1}{^{2\pi}}
\int\limits_{[0,T]\times\BZ}V_\epsilon^*H\,,
\qqq
see (\ref{deg}), \,i.e. as \,the homotopy invariant of the map 
$\,V_\epsilon:\mathbb R/(T\mathbb Z)\times\BZ\rightarrow U(N)$. For two gaps with
$\,-\frac{2\pi}{T}<\epsilon_1<\epsilon_2<0$,
\qq
V_{\epsilon_2}(t,k)\,=\,V_{\epsilon_1}(t,k)\,\ee^{\frac{2\pi\ii t}{T}
P_{\epsilon_1,\epsilon_2}(k)},
\qqq
due to (\ref{2epsilons}). As 
$\,{\rm deg}(V)\,$ is additive under the multiplication of maps, it follows 
from relation (\ref{deg}) that
\qq
W_{\epsilon_2}-W_{\epsilon_1}\,=\,c_{\epsilon_1,\epsilon_2}\,,
\label{2Ws}
\qqq
where $\,c_{\epsilon_1,\epsilon_2}\,$ is \,the $\,1^{\rm st}$ Chern number of 
the vector bundle $\,\CE_{\epsilon_1,\epsilon_2}\,$ with fibers 
$\,P_{\epsilon_1,\epsilon_2}(k)\,\mathbb C^N$,
\,i.e. spanned by the eigenstates of $\,\hat U(t,k)\,$ with the quasi-energies
between the two gaps. In particular, we may have the bundles of states between
two gaps topologically trivial with $\,c_{\epsilon_1,\epsilon_2}=0\,$ 
but the Floquet systems still topologically nontrivial
with $\,W_{\epsilon_1}=W_{\epsilon_2}\not=0$. 
\vskip 0.2cm

In \cite{CDFG,CDFGT}, we considered the time-reversal invariant gapped
Floquet systems with
\qq
H_{_\CC}(-t)\,=\,\theta\,H_{_\CC}(t)\,\theta^{-1}
\qquad{\rm or,\ \ equivalently,}\qquad
\hat H(-t,-k)=\theta\,\hat H(t,k)\,\theta^{-1}\,.
\qqq
In this case,
\qq
\hat U(-t,-k)\,=\,\theta\,\hat U(t,k)\,\theta^{-1}\,,
\qquad H^{\rm eff}_\epsilon(-k)=\theta H^{\rm eff}_\epsilon(k)\,\theta^{-1}
\qqq
and 
\qq
V_\epsilon(T-t,-k)\,=\,V_\epsilon(-t,-k)\,=\,\theta\,V_\epsilon(t,k)\,\theta^{-1}\,.
\qqq
The latter symmetry property implies that in $2d$ the index 
$\,W_\epsilon\,$ of \cite{RLBL} vanishes. \,Instead, we introduced an 
index $\,K_\epsilon\in\mathbb Z_2\,$ defined by the relation
\qq
(-1)^{K_\epsilon}\,=\,\CK(V_\epsilon)\,,
\label{2dKepsilon}
\qqq
where $\,\CK(V_\epsilon)\,$ is the $3d$ index with values in $\,\pm1\,$
defined in  Sec.\,\ref{subsec:3d_ind} for equivariant maps. Index 
$\,K_\epsilon\,$ is a topological invariant of time reversal symmetric 
gapped Floquet systems in $2d$. \,In the case of two gaps with 
$\,-\frac{2\pi}{T}<\epsilon_1<\epsilon_2<0$,
\qq
K_{\epsilon_2}-K_{\epsilon_1}\,=\,\KM_{\epsilon_1,\epsilon_2}\,,
\label{KKKM}
\qqq
where $\,\KM_{\epsilon_1,\epsilon_2}\in\mathbb Z_2\,$ is the Kane-Mele index of 
the vector bundle $\,\CE_{\epsilon_1,\epsilon_2}\,$ with fibers 
$\,P_{\epsilon_1,\epsilon_2}(k)\,\mathbb C^N$. $\KM_{\epsilon_1,\epsilon_2}$ is well 
defined since $\,P_{\epsilon_1,\epsilon_2}(-k)=\theta\,P_{\epsilon_1,\epsilon_2}(k)
\,\theta^{-1}$. \,The proof of relation (\ref{KKKM}) uses the fact that
\qq
\CK(V_{\epsilon_2})\,=\,\CK(V_{\epsilon_1}\Phi)\,=\,
\CK(V_{\epsilon_1})\,\CK(\Phi)\,,
\qqq
for $\,\Phi(t,k)=\ee^{\frac{2\pi\ii t}{T}P(k)}\,$
(the $3d$ index is not multiplicative in general, but
in this case it is) and the relation (\ref{KMWZ}).
\vskip 0.2cm

Finally, for gapped $3d$ Floquet systems, we may define $\,6\,$
week indices $\,K_\epsilon({i,a})\in\mathbb Z_2\,$ for $i=1,2,3$ and 
$a=0,\pi$ from 6 $\,2d$ sub-tori $\,\mathbb T^2({i,a})\,$ in the 
$3d$ Brillouin torus $\,\BZ\cong\mathbb T^3$, \,see Fig.\,\ref{fig:torus3d}, 
and one strong index $\,K^s_\epsilon\in\mathbb Z_2\,$ expressed by the $3d$ 
index for the equivariant map $\,\BZ\ni k\longmapsto V_\epsilon(\frac{T}{2},k)$,
\qq
(-1)^{K^s_{\epsilon}}\,=\,\CK(V_\epsilon|_{t=\frac{T}{2}})\,.
\qqq
Again for two gaps with $\,-\frac{2\pi}{T}<\epsilon_1<\epsilon_2<0$,
\qq
&&(-1)^{K^s_{\epsilon_2}}\,=\,\CK(V_{\epsilon_2}|_{t=\frac{T}{2}})
=\CK(V_{\epsilon_1}|_{t=\frac{T}{2}}
(1-2P_{\epsilon_1,\epsilon_2}))\cr
&&=\,\CK(V_{\epsilon_1}|_{t=\frac{T}{2}})\,\CK(1-2P_{\epsilon_1,\epsilon_2})\,
=\,(-1)^{K^s_{\epsilon_1}}\,(-1)^{\KM^s_{\epsilon_1,\epsilon_2}}\,,
\qqq
where we used the multiplicativity of the $3d$ index holding
for maps with no winding of determinants and the relation 
(\ref{KMsannoun}) of Sec.\,\ref{subsec:3d_KM_ind}. Thus
\qq
K^s_{\epsilon_2}-K^s_{\epsilon_1}\,=\,\KM^s_{\epsilon_1,\epsilon_2}\,.
\qqq
similarly as in $2d$, \,see (\ref{KKKM}). One also has
\qq
K^s_{\epsilon}\,=\,K_\epsilon({i,\pi})-K_\epsilon({i,0})
\label{KsKK}
\qqq
for each $\,i=1,2,3$, \,similarly as for the $3d$ Kane-Mele indices, see 
(\ref{KMsKMKM}). For $i=1$ this is proven by using the fundamental
domain $\,F_R=\BZ_+\,$ for the involution $\,\rho:k\mapsto-k\,$ of $\,\BZ\,$
and observing that the relation
\qq
\int\limits_{\partial\widetilde{\BZ}_+}V_\epsilon^*H\,=\,0
\qqq
for $\,\widetilde{\BZ}_+=[0,\frac{1}{2}]\times\BZ_+\,$ implies that
\qq
\int\limits_{\BZ_+}(V_\epsilon|_{t=\frac{T}{2}})^*H\,=\,\int_{[0.\frac{1}{2}]\times T^{1,\pi}}
\hspace{-0.2cm}V_\epsilon^*H\ -\int_{[0.\frac{1}{2}]\times T^{1,0}}\hspace{-0.2cm}V_\epsilon^*H  
\qqq
so that
\qq
\CK(V_\epsilon|_{t=\frac{T}{2}})\,=\,\frac{\CK(V_\epsilon|_{k_1=\pi})}{\CK(V_\epsilon|_{k_1=0})}\,.
\qqq
The latter equality, together with the $2d$ relation (\ref{2dKepsilon}), 
implies (\ref{KsKK}) for $\,i=1$. \,The proof for $\,i=2,3\,$ goes the 
same way by choosing the fundamental domains for $\,\rho\,$ corresponding
to the restrictions $\,k_i\in[0,\pi]$.

\subsection{Bulk-edge correspondence}
\label{subsec:bulk_edge}

\begin{figure}[h]
\begin{center}
\leavevmode
\hspace{-0.3cm}
{%
      \begin{minipage}{0.43\textwidth}
        \includegraphics[width=3.5cm,height=3.1cm]{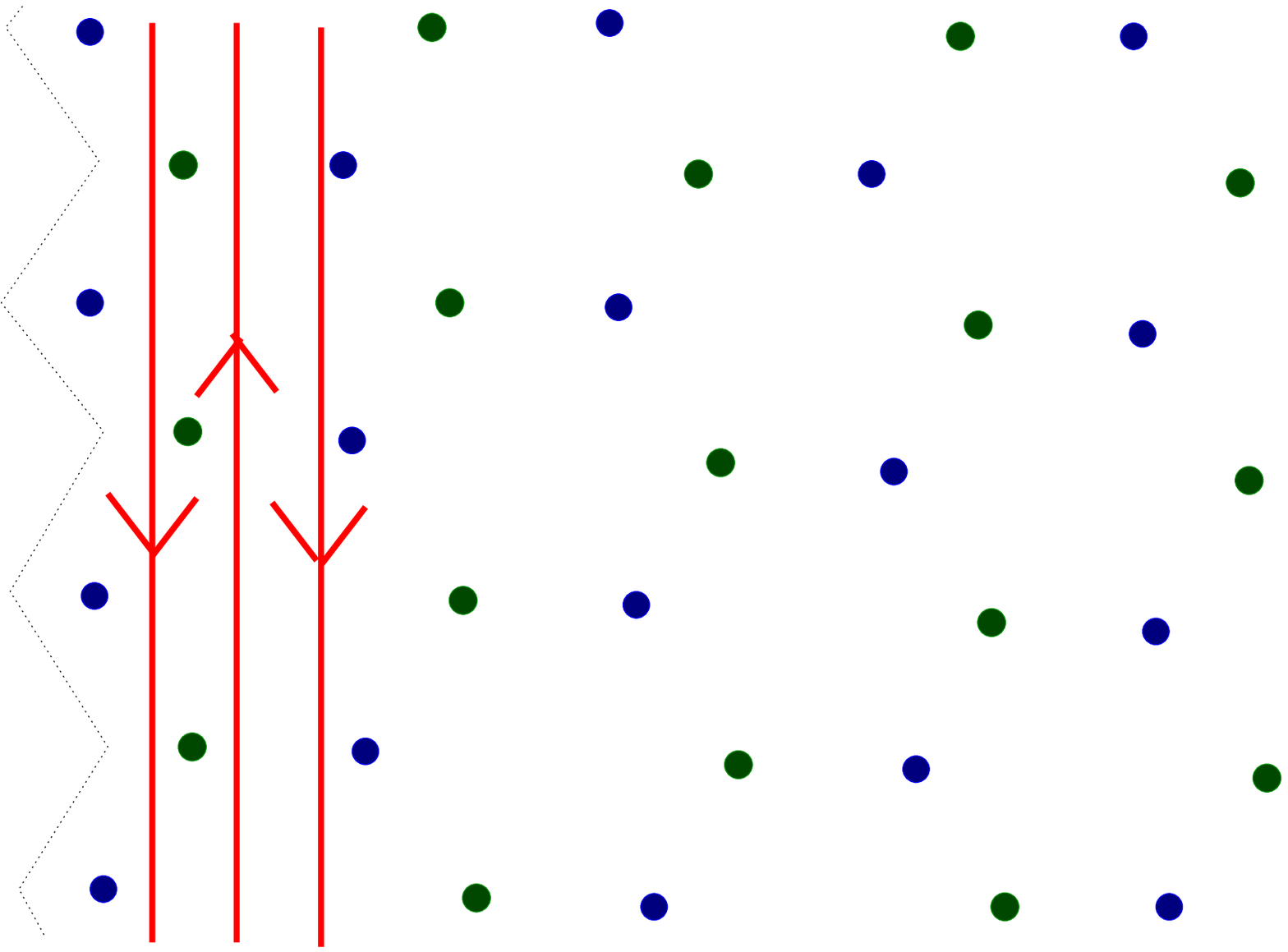}\\
        \vspace{-0.8cm} \strut
        \end{minipage}}
    \hspace*{-3cm}
{%
      \begin{minipage}{0.43\textwidth}
        \includegraphics[width=3.8cm,height=3.7cm]{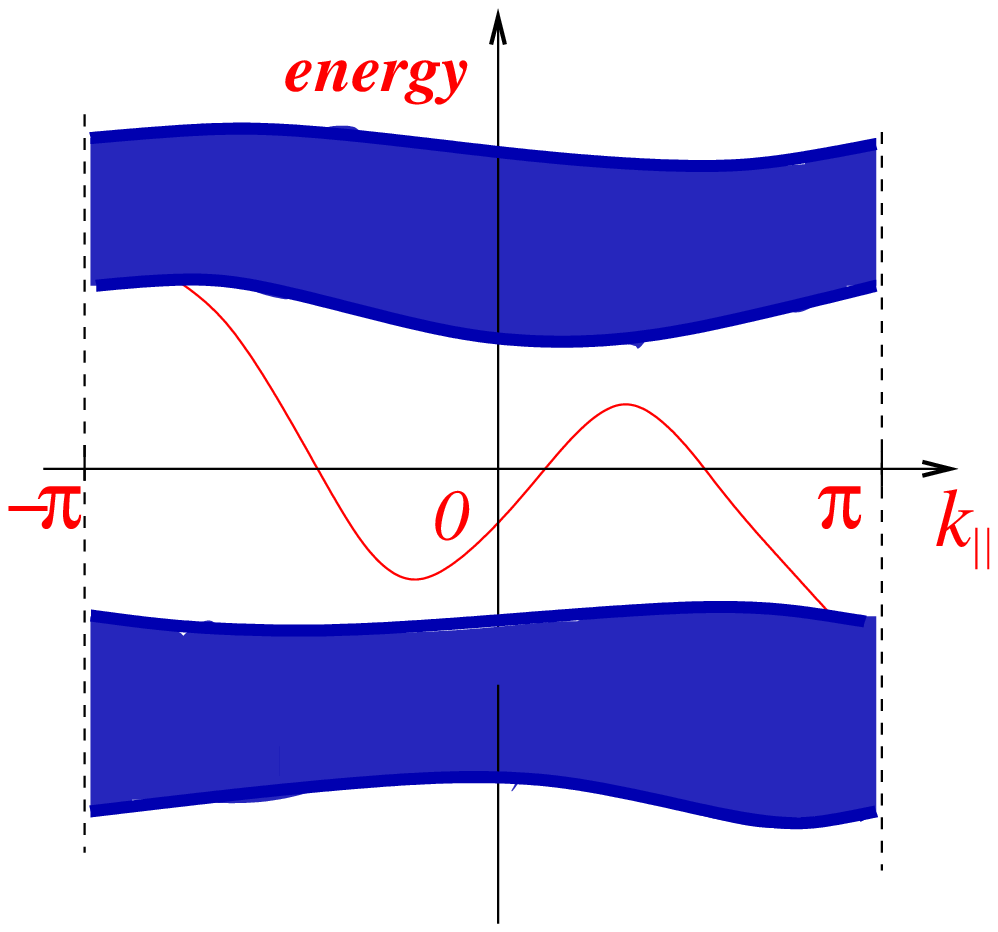}\\
      \vspace{-0.8cm} \strut
        \end{minipage}}
\end{center}
\caption{The edge current (left) and the spectrum of
Bloch Hamiltonians for the half-space geometry (right)}
\label{fig:edge1}
\vskip 0.1cm
\end{figure}

\noindent Physically, the most interesting aspect of topological phases 
of insulators, the mean of their detection, and the essential feature 
for present and future application, is that, in finite geometry, they may carry 
currents localized on the boundary of the samples that are protected against 
sufficiently weak disorder or interactions. In particular, in the half-space 
geometry, such edge currents are localized near the boundary hyperplane, \,see 
the left part of \,Fig.\,\ref{fig:edge1}. The massless modes carrying the edge 
currents can be detected in the spectrum of the half-crystal Hamiltonian 
Fourier transformed in the directions parallel to the boundary hyperplane. 
They correspond to the intersections with the Fermi energy level of the 
energy bands that appear in such boundary Hamiltonians within the gap 
of the original crystalline Hamiltonian, see the right part of 
\,Fig.\,\ref{fig:edge1}.  
\vskip 0.2cm

\begin{figure}[h]
\vskip -0.0cm
\leavevmode
\begin{center}
\hspace*{-0.1cm}
\includegraphics*[width=7.6cm,height=3.6cm,angle=0]{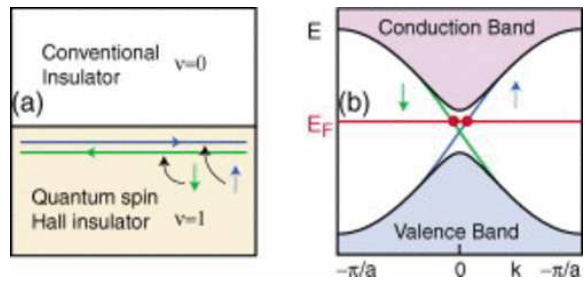}\\
\end{center}
\vskip -0.2cm
\caption{Pairs of chiral edge currents (left) and the typical spectrum of
time reversal symmetric half-space Bloch Hamiltonians (right)}
\label{fig:edge2}
\end{figure}

\noindent In particular, in a $2d$ Chern insulator, the $1^{\rm st}$ Chern 
number $\,c_1$ 
of the infinite-crystal valence states bundle counts with chirality
(i.e. direction) the massless modes of the half-crystal Bloch Hamiltonians 
with energies close to the Fermi energy $\,\epsilon_F\,$ ($=0\,$ in 
Fig.\,\ref{fig:edge1} corresponding to $\,c_1=1$), \,see e.g. \cite{QWZ}. 
For $2d$ time reversal invariant insulators, the Kane-Mele invariant $\,\KM\,$
counts the parity of the number of Kramers pairs of massless edge modes 
of opposite chirality, \,see Fig.\,\ref{fig:edge2} taken from \cite{HK} that 
corresponds to $\,\KM=1$. 

\noindent The bulk-edge correspondence in this case was proven 
rigorously in \cite{GrafPorta}. Similarly, the strong Kane-Mele $3d$ invariant
should count the parity of the number of Dirac cones intersecting the Fermi 
energy inside the bulk gap.

\begin{figure}[h!]
\vskip -0.8cm
\leavevmode
\begin{center}
\hspace*{-1.2cm}
\includegraphics*[width=4.9cm,height=8.9cm,angle=-90]{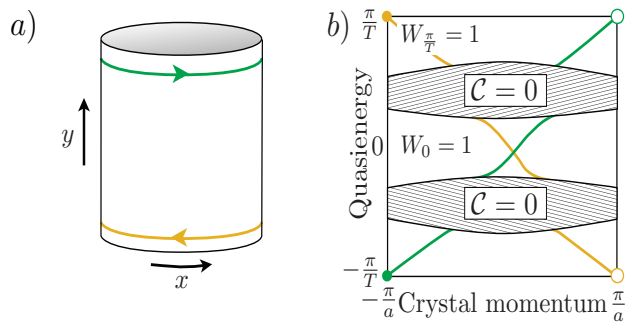}\\
\end{center}
\vskip -0.1cm
\caption{Possible quasi-energy spectrum of a gapped Floquet 
system in the strip geometry}
\label{fig:nontrWW}
\end{figure}

It was shown in \cite{RLBL} that for the gapped Floquet systems 
the invariant $\,W_\epsilon\,$ enumerates with chirality the edge massless
modes appearing in the quasi-energy gap around $\,\epsilon\,$ even
in situations when the $1^{\rm st}$ Chern numbers of quasi-energy bands 
vanish, as in Fig.\,\ref{fig:nontrWW} 
\,extracted from \cite{TBRRL}.
\vskip 0.1cm

\vskip 0.2cm

\begin{figure}[h]
\leavevmode
\begin{center}
\vskip 0.1cm
\hspace*{-0.4cm}
\includegraphics*[width=8.4cm,height=5cm,angle=0]{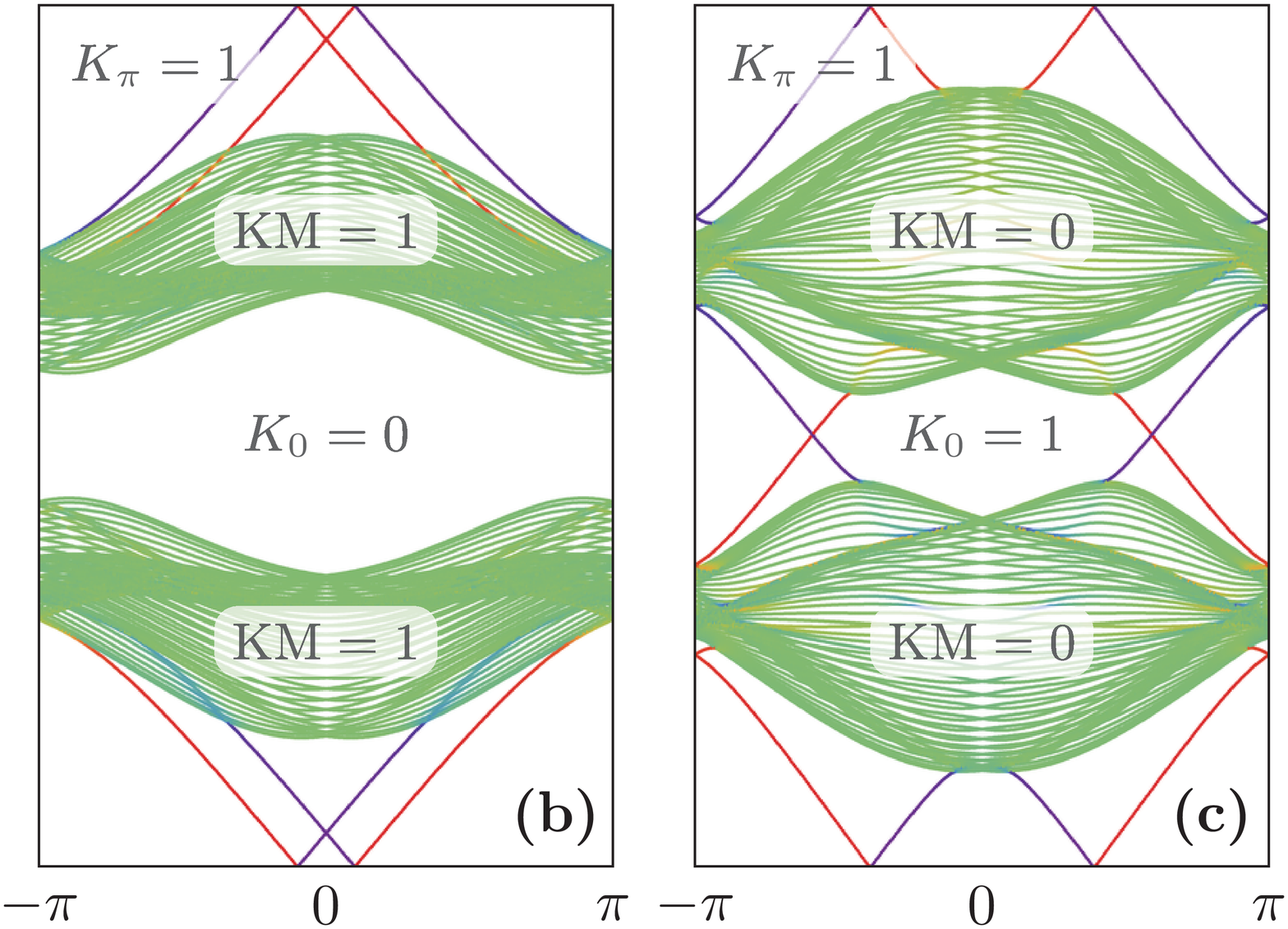}\\
\end{center}
\vskip -0.3cm
\caption{Quasi-energy spectra of a simple $2d$ Floquet system with time 
reversal symmetry in the strip geometry}
\label{fig:Fledgespec}
\vskip -0.1cm
\end{figure}

Similarly, for time reversal invariant $2d$ gapped Floquet systems, 
invariant $\,K_\epsilon\,$ should count modulo two the number of Kramers pairs
of massless edge modes in the bulk quasi-energy gap around $\,\epsilon$,
\,as confirmed by simulations reported in \cite{CDFG} of a simple periodically
driven model in a strip geometry, \,see Fig.\,\ref{fig:Fledgespec}.  

\subsection{Conclusions} 

\noindent We discussed how $\,\Theta$-equivariant structures 
on a bundle gerbe $\,\CG\,$ over manifold $\,M\,$ equipped with involution
$\,\Theta\,$ may be used to fix the square root of the WZ amplitudes 
of maps from $2d$ surfaces with orientation preserving 
involutions to $\,M$. \,Using such square roots, we
defined an index $\,\CK(\Phi)=\pm1\,$ of equivariant maps 
$\,\Phi\,$ from a $3d$ manifolds equipped with an orientation 
reversing involutions to $\,M$. Such index was applied to the
situation when $\,M=U(N)$, $\,\CG\,$ is the basic gerbe on $\,U(N)\,$
with the bi-invariant curvature, and $\,\Theta\,$ is given by the 
adjoint action of an anti-unitary transformation of $\,\mathbb C^N\,$ 
that squares to $\,-I$. The problem with the absence of a 
$\Theta$-equivariant structure on $\,\CG\,$ was solved by passing
to the double cover of $\,U(N)$. \,We discussed how index $\,\CK(\Phi)\,$ 
may be used to express the Kane-Mele $\,\mathbb Z_2$-valued
invariant of $2d$ and $3d$ time-reversal invariant insulators and
to obtain its generalization to periodically driven crystalline systems.
\vskip 0.1cm   
       
The approach discussed here, based on the geometry of bundle gerbes,
may be also extended to topological insulators in other symmetry classes
that lead to torsion invariants, and, potentially, to insulators with more
crystalline symmetries. It should be also useful to describe the bulk-edge 
correspondence, but that remains to be done constituting the main challenge 
for this line of thought. A possible extension of the present approach to 
non-commutative geometry should be investigated. 
\vskip 0.6cm

\noindent{\bf Acknowledgements.} \,I thank my collaborators: D. Carpentier,
P. Delplace, M. Fruchart and C. Tauber for many discussions that helped
crystallize the ideas presented in these lectures. I am grateful to
J. Kellendonk for his lectures about the $C^*$-algebraic approach to 
topological insulators, where similar ideas may find their natural extension. 
Special thanks are due to A. Alekseev for the invitation to give a mini-course 
on the topic of these notes in Geneva.


\end{document}